\documentclass[12pt]{article}
\usepackage{amsmath,amssymb,multirow,jheppub2,graphicx}
\usepackage{centernot}
\usepackage{mciteplus}
\usepackage{bbm}
\usepackage{color}
\allowdisplaybreaks[1]
\usepackage{tikz}
 \tikzset{node distance=2cm, auto}

\def\tr{\text{tr}}

\def\tfrac#1#2{{\textstyle{\frac{#1}{#2}}}}

\def\hat{\widehat}
\def\bar{\overline}

\def\be{\begin{equation}}
\def\ee{\end{equation}}
\def\Z{{\mathbb Z}}

\usepackage{verbatim}   
\usepackage[all]{xy}
\usepackage{bm}  
\def\Dslash{{\rlap{\raise 1pt \hbox{$\>/$}}D}}
\def\Pslash{{\rlap{\raise  1pt \hbox{$\>/$}}\,\partial}}

\def\Th#1#2{\vartheta{\tiny\begin{bmatrix}
{#1}\\
{#2}
\end{bmatrix}}}

\begin{document}
\begin{flushright}
FTPI-MINN-18-20, UMN-TH-3803/18
\end{flushright}

\title{Bose-Fermi cancellations without supersymmetry} 
 
\author[a,d]{Aleksey Cherman\,} 
\author[b,d]{\!\!, Mikhail Shifman\,} 
\author[c,d]{\!\!,  Mithat \"Unsal\,}
\emailAdd{aleksey.cherman.physics@gmail.com}
\emailAdd{shifman@umn.edu}
\emailAdd{unsal.mithat@gmail.com}
\affiliation[a]{Department pf Physics, University of Minnesota, Minneapolis MN, USA}
\affiliation[b]{Fine Theoretical Physics Institute, University of Minnesota, Minneapolis MN, USA}
\affiliation[c]{Department of Physics, North Carolina State University,  Raleigh, NC 27695, USA}
\affiliation[d]{Kavli Institute for Theoretical Physics University of California, Santa Barbara, CA 93106}

\abstract{We show that adjoint QCD features very strong Bose-Fermi cancellations in the large $N$ limit, despite the fact that it is manifestly non-supersymmetric.  The difference between the bosonic and fermionic densities of states in large $N$ adjoint QCD turns out to have a `two-dimensional'  scaling $\sim \exp{(\sqrt{\ell E})}$ for large energies $E$ in finite spatial volume, where $\ell$ is a length scale associated with the curvature of the spatial manifold.   In particular, all Hagedorn growth cancels, and so does the growth $\exp{(V^{1/4} E^{3/4})}$ expected in a standard local 4d theory in spatial volume $V$.  In these ways, large $N$ adjoint QCD, a manifestly non-supersymmetric theory, acts similarly to supersymmetric theories.  We also show that at large $N$, the vacuum energy of multi-flavor adjoint QCD is non-negative and exponentially small compared to the UV cutoff with several natural regulators. }

\maketitle


\section{Introduction} 
The goal of this paper is to discuss relations between bosonic and fermionic excitations in four-dimensional adjoint QCD.  
Despite the manifest lack of supersymmetry in adjoint QCD with $n_f>1$, these relations turn out to be surprisingly powerful.  
In several ways these relations turn out to be as  powerful as the Bose-Fermi relations in supersymmetric QFTs!

To probe relations between bosonic and fermionic states, we will mostly consider a $(-1)^F$-graded grand-canonical partition function $\tilde{Z}(L)$ and the related grand-canonical $(-1)^F$-graded density of states $\tilde{\rho}(E)$:
\begin{align}
\tilde{Z}(L)  = \tr (-1)^F e^{-L H} = \int dE \, \tilde{\rho}(E) e^{-L E}
\end{align}
Here $\tilde{\rho}(E) = \rho_B(E) - \rho_F(E)$, and $\rho_B(E)$ and $\rho_F(E)$ are the bosonic and fermionic densities of states as a function of energy $E$.  

 In four-dimensional supersymmetric quantum field theories (SUSY QFTs), the energies of bosonic and fermionic states are tightly correlated by definition.  In flat space, bosonic and fermionic finite-energy excitations come in degenerate pairs, and (at least when the spectrum is discrete) $\tilde{\rho}(E)$ vanishes for energies $E>0$, and $\tilde{Z}$ becomes the Witten index\cite{Witten:1982df}.  If space is taken to be a compact curved manifold, then in a SUSY QFT \cite{DiPietro:2014bca} 
\begin{subequations}
\begin{align}
\log \tilde{\rho}(E) &\sim  \sqrt{\ell E}, \\ 
\log \tilde{Z}(L) &\sim \frac{\ell}{L} 
\end{align} 
\label{eq:SUSYscaling}
\end{subequations}
where $\sim$ indicates the scaling for large $E$ and small $L$ respectively,  $\ell$ is a length scale characterizing the spatial manifold $M$, $\ell \equiv \int d^3{x} \sqrt{g}\, \mathcal{R}$, and $g$ and $\mathcal{R}$ are the metric and Ricci scalar curvature of $M$.\footnote{\label{footnote:ell_definition} It can be helpful to write $\ell = V/R^2$, where $V$ is the volume of $M$ and $R$ is its volume-averaged radius of curvature.  For example, if $M = T^3$, which is flat, then $\ell = 0$, but if $M = S^3$ with radius $r$ then $\ell \sim r$. }

In generic non-supersymmetric 4D QFTs, on the other hand,  one expects
\begin{align}
\textrm{no SUSY} \;\; \Longrightarrow \;\; 
\begin{cases}
\;\;\log \tilde{\rho}(E)   \sim V^{1/4} E^{3/4} \\
\;\;\log \tilde{Z}(L)  \sim V/L^3, 
\end{cases}
 \label{eq:nonSUSYscaling}
\end{align}
These scaling relations follow from the expectation that the partition function should have an extensive dependence on the spatial volume $V$  in the absence of high-energy Bose-Fermi cancellations.  Indeed, roughly speaking the coefficient of $V/L^3$ counts the difference between the number of bosonic and fermionic degrees of freedom at short distances. 
Its value can be related to the standard quartically-divergent contribution to the cosmological constant, as we discuss in Section~\ref{sec:CC}.   Of course, SUSY implies that the Bose-Fermi degree-of-freedom mismatch underlying Eq.~\eqref{eq:nonSUSYscaling} vanish, and, relatedly, also implies that quartic divergences in the contributions to the vacuum energy must vanish, leading to Eq.~\eqref{eq:SUSYscaling}.    

The main goal of this paper is to show that there are some manifestly non-supersymmetric QFTs which  manage to satisfy Eq.~\eqref{eq:SUSYscaling}. In particular, we find that Eq.~\eqref{eq:SUSYscaling} applies to $U(N)$ adjoint QCD coupled to $1 \le n_f < 6$ massless adjoint Majorana fermions in the 't Hooft large $N$ limit.   This family of theories is asymptotically free, with a strong scale $\Lambda$.\footnote{For some values of $n_f$, especially $n_f = 5$ and most likely also $n_f = 4$~\cite{Catterall:2007yx,Catterall:2008qk,Hietanen:2008mr,Hietanen:2009az,DelDebbio:2008zf,Poppitz:2009uq,Poppitz:2009tw,DelDebbio:2010hx,DeGrand:2011qd,Bursa:2011ru,Shifman:2013yca,Bergner:2016hip}, the theory is believed to be in an infrared-conformal phase.  The lower boundary of the conformal window is not known.  For theories in the conformal window one can interpret $\Lambda$ as the scale at which the gauge coupling saturates to its infrared-fixed-point value. }   
We choose to define $U(N)$ adjoint QCD to include $n_f$ massless free Majorana fermions in addition to a Maxwell field.    Then if $n_f = 1$, adjoint QCD reduces to $\mathcal{N}=1$ pure super-Yang-Mills theory, and the fact that Eq.~\eqref{eq:SUSYscaling} is satisfied when $n_f=1$ is not surprising.   But for $n_f > 1$, the number of massless microscopic bosonic degrees of freedom $\sim N^2$ is  smaller than the number of massless fermionic degrees of freedom $\sim N^2 n_f$, so  there is no supersymmetry to begin, and there is naively no reason that Eq.~\eqref{eq:SUSYscaling} should hold.

\begin{figure}[t]
\centering
\includegraphics[width=.8\textwidth]{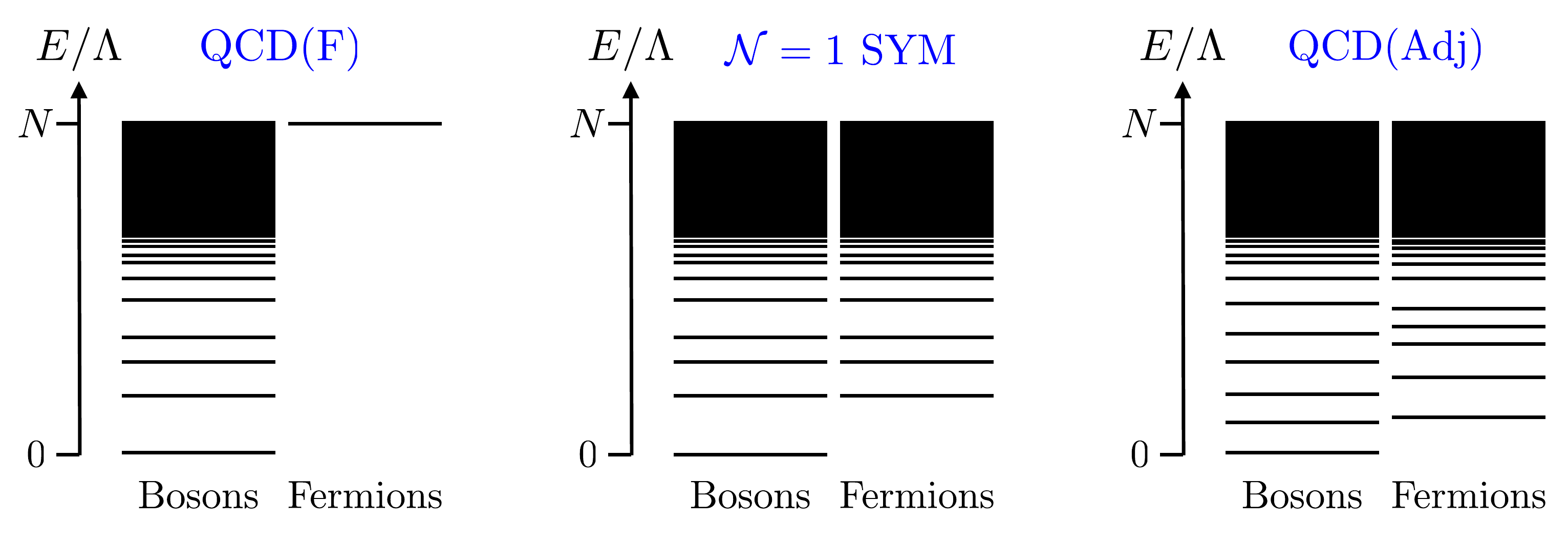}
\caption
    {%
    Cartoons illustrations of the large $N$ bosonic and fermionic Hilbert spaces of color-singlet states in QCD with fundamental fermions (left), $\mathcal{N}=1$ super-Yang-Mills (middle), and multi-flavor adjoint QCD (right).  The are no light fermionic states in QCD(F), so no Bose-Fermi correlations are possible.  In $\mathcal{N}=1$ SYM the non-zero energy states come in Bose-Fermi pairs due to supersymmetry.  In QCD(Adj) there is no mode-by-mode Bose-Fermi pairing and no supersymmetry, but at large $N$ the spectrum nevertheless enjoys Bose-Fermi correlations, with some consequences as powerful as the ones that follow  from supersymmetry. 
    }
\label{fig:hilbert_space}
\end{figure}

Nevertheless, we find that at small $L\Lambda$, the graded partition function scales as 
\begin{align}
\log \tilde{Z}(L) \sim \frac{a}{N^2} \frac{V}{L^3} + b \frac{\ell}{L} \,,\quad \textrm{$U(N)$ adjoint QCD} 
\label{eq:UNresult}
\end{align}
for any $1 < n_f < 6$.  Here $a$ and $b$ are dimensionless parameters which are independent of $N$ in the large $N$ limit, and have a logarithmic dependence on $L$.  
Note that the first term is suppressed by $1/N^4$ relative to its naive $\sim N^2$ scaling.  
If one takes the large $N$ limit with all other parameters held fixed, the first term vanishes, and we recover Eq.~\eqref{eq:SUSYscaling}.  
Consequently, the $(-1)^F$-graded partition function of four-dimensional  large $N$ adjoint QCD behaves as if were the partition function of a \emph{two-dimensional} theory.  
The general observation that large $N$ adjoint QCD must have strong Bose-Fermi relations was first made in Ref.~\cite{Basar:2013sza}, while the observation that these cancellations can be so strong as to lead to Eq.~\eqref{eq:SUSYscaling} was made in Ref.~\cite{Basar:2014jua,Basar:2014hda,Basar:2015asd,Basar:2015xda,Basar:2014mha,Cherman:2015rpc} in the context of a one-loop analysis.  Our results here generalize these earlier works and promote Eq.~\eqref{eq:SUSYscaling} to an exact statement about adjoint QCD.

How is such a thing possible without supersymmetry, given the obvious mismatch in the number of bosonic and fermionic degrees of freedom in adjoint QCD?  The answer is tied up with two special features of adjoint QCD.  The first special feature is that adjoint QCD has \emph{both} bosonic and fermionic color-singlet excitations with energies $\sim N^0$.  Among theories with fermionic matter fields in a single color representation of $SU(N)$,  the only way to achieve this is with fermions in a real representation, and the only real representation $SU(N)$ is the adjoint representation.  In QCD with e.g. fundamental fermions, the lightest fermionic states have energies $\sim N$ when $N$ is odd, and there are no fermionic states at all when $N$ is even.  This feature is illustrated in Fig.~\ref{fig:hilbert_space}.

The second special feature of adjoint QCD is that if it is compactified on a circle of size $L$ with periodic boundary conditions, it stays in the confined phase for all $L$, even if $L$ is small, and has a smooth dependence on $L$, see e.g. \cite{Kovtun:2007py,Unsal:2007jx,Unsal:2008eg,Unsal:2008ch,Unsal:2007vu, Argyres:2012ka, Anber:2015kea, Anber:2015wha, Anber:2013doa, Anber:2014lba,Bringoltz:2009kb,  Bringoltz:2011by, Azeyanagi:2010ne, Cossu:2009sq,Cossu:2013nla,Cossu:2013ora, Hietanen:2009ex,Hietanen:2010fx}.  Note that Eq.~\eqref{eq:UNresult} is obtained precisely at small $L\Lambda$, where adjoint QCD is in the confined phase.   Relatedly, in adjoint QCD, the $L$ dependence of appropriate observables disappears at large $N$ in a phenomenon called large $N$ volume independence\cite{Kovtun:2007py}, see also e.g.~\cite{Eguchi:1982nm,Bhanot:1982sh,GonzalezArroyo:1982ub,GonzalezArroyo:1982hz,GonzalezArroyo:2010ss,Gonzalez-Arroyo:2013bta,Perez:2013dra,Perez:2014sqa} for important related work.   These two features are tied to the $\mathbb{Z}_N$ center symmetry of adjoint QCD. The Euclidean path integral associated to compactification on a circle with periodic boundary conditions calculates precisely the $(-1)^F$ graded partition function we are discussing.  One can thus think about the contributions to $\log \tilde{Z}$ at small $L\Lambda$ in two ways:  either as a $(-1)^F$-graded sum of contributions of colorless hadronic states, or as a (gauge-invariant) $(-1)^F$-graded sum of colored gluon and adjoint quark contributions.   

Introducing a $(-1)^F$ grading makes adjoint QCD remain confining at small $L$, in the sense that center symmetry is not spontaneously broken.  The reason this is relevant is because, thanks to the presence of a center-symmetric Polyakov loop expectation value, the contributions of the quarks and gluons to the partition function come with phases which are $N$-th roots of $1$.   These phases induce extreme destructive interference in the sum over colors, suppressing the coefficient of $L^{-3}$ by four powers of $N$ relative to the naively expected $N^2 L^{-3}$ term, as advertised in Eq.~\eqref{eq:UNresult}.    This phenomenon is explained in Section~\ref{sec:quick_derivation}.
From the hadronic perspective, this means that the energies and distributions of bosonic and fermionic hadrons are such that they manage to cancel each other to extreme accuracy in $\log \tilde{Z}$ despite the absence of energy-level-by-energy-level cancellations in this non-supersymmetric QFT.  

These cancellations require a subtle ``spectral conspiracy" or ``emergent large $N$ symmetry" which is tied up with \emph{both} confinement and  the large $N$ limit.   The ``emergent large $N$ symmetry" terminology was first used in  Ref.~\cite{Basar:2013sza}, and we feel that the power of the Bose-Fermi relations in adjoint QCD justifies this term.  But we emphasize that we are not dealing with any standard symmetry, because there are no level-by-level Bose-Fermi cancellations in adjoint QCD;  the cancellations non-trivially involve summing over the whole spectrum, as discussed in \cite{Basar:2013sza,Basar:2014jua} and in this paper.  That is why we generally use the less prejudicial term ``spectral conspiracy" in this paper.

The fact that the large $N$ limit is necessary for the cancellations is naively rather surprising, because one might have thought it would make matters worse rather than better.  In confining large $N$ gauge theories, the density of states with energies $E \sim N^0$ scales as 
\begin{align} 
\rho(E) \sim e^{L_H E} + \cdots
\end{align}
 for some length scale $L_H$, leading to `Hagedorn singularities' in the partition function.  Note that the $\cdots$ terms generically include an infinite number of smaller but still exponentially growing terms, as we discuss in more detail in Section~\ref{sec:spectral_conspiracy}.  The fact that adjoint QCD remains confining at small $L$,
 with no phase transition between small and large $L$,  means that all such exponential growth must cancel between the bosons and fermions, as emphasized in Refs.~\cite{Basar:2013sza} and \cite{Basar:2014jua}.  Such Hagedorn cancellations are very difficult to achieve.   The point we emphasize here is that the cancellations are even more severe than expected in Ref.~\cite{Basar:2013sza}:  they are so precise that not even a single 4d-particle's worth of degree of freedom is left over, despite the absence of supersymmetry.

To establish Eq.~\eqref{eq:UNresult}, our basic tool is to study compactification of adjoint QCD on a circle of circumference $L$ with periodic boundary conditions for the fermions.  In Section~\ref{sec:volume_independence} we show that confinement is indeed present at small $L\Lambda$ at large $N$ in adjoint QCD, filling a small gap in the discussion of Ref.~\cite{Kovtun:2007py} along the way.   We then analyze the implications of known results on large $N$ volume independence on the partition function, and show that they imply a version of Eq.~\eqref{eq:SUSYscaling} for $\mathbb{R}^3 \times S^1$.   This gives a simple but not especially explicit demonstration of our main result.  Section~\ref{sec:cancellations} contains a more direct analysis of the graded partition function, and shows that the preservation of center symmetry at small $L \Lambda$ leads to  Eq.~\eqref{eq:SUSYscaling}.    In Section~\ref{sec:deformations} we consider deformations of adjoint QCD obtained by turning on quark masses or introducing some extra grading by the flavor symmetries into the partition function to explore the class of theories and symmetry gradings that can produce Eq.~\eqref{eq:SUSYscaling}.\footnote{For example, when  $n_f = 2$ not only $(-1)^F $ is well-defined, but so is the fermion number $F$ per se\cite{Shifman:2013yca}. So one can introduce gradings by $e^{i\alpha F}$ rather than just $e^{i \pi F}$.}   As a side benefit, by a method similar to Ref.~\cite{Cherman:2016hcd}, we construct compactifications of adjoint QCD that have a smooth dependence on $L$, regardless of the possible fate of chiral symmetry breaking in adjoint QCD on $\mathbb{R}^4$.\footnote{For recent discussions of chiral symmetry breaking, or its possible absence, in $N=2$ adjoint QCD see  e.g.~\cite{Athenodorou:2014eua,Bergner:2016hip,Bergner:2017gzw,Anber:2018tcj,Cordova:2018acb,Bi:2018xvr}.  Reference ~\cite{Armoni:2004uu} discusses the expectations for $N>2$, as well as the relation of adjoint QCD to a bona-fide large $N$ limit of standard QCD via the `orientifold large $N$ equivalence between adjoint QCD and QCD with two-index antisymmetric representation quarks. }  Section~\ref{sec:discussion} places our results in context.  There we discuss the connections between our work here and earlier discussions of Bose-Fermi cancellations in adjoint QCD\cite{Basar:2013sza,Basar:2014jua}, and discuss a striking parallel between our results and the notion of misaligned supersymmetry in string theory\cite{Kutasov:1990sv,Dienes:1994np,Dienes:1994jt,Dienes:1995pm}.  Finally, we explain the implications of our results for the vacuum energy $\langle E \rangle$ of multi-flavor large $N$ adjoint QCD, and argue that  $\langle E \rangle$ is non-negative and exponentially small in units of the UV cutoff, with several natural choices for UV regulators.

\section{Large $N$ volume independence and its implications}
\label{sec:volume_independence}
In this section we explain how massless adjoint QCD manages to remain in the confined phase when compactified on a small circle, and then show that putting this result together with known properties of large $N$ volume independence leads to our main result, Eq.~\eqref{eq:UNresult}.

\subsection{Confinement at small $L\Lambda$}
\label{sec:small_L_confinement}

Reference~\cite{Kovtun:2007py} showed that adjoint QCD with massless quarks has unbroken center symmetry when it is compactified on $\mathbb{R}^3 \times S^1$ with sufficiently small $S^1$ sizes $L$, so long as periodic boundary conditions are used for all of the fields (including the quarks).  At large $L$ there is evidence from lattice simulations that the theory is in a center-symmetric phase, see e.g.~\cite{Bringoltz:2009kb,Bringoltz:2011by,Azeyanagi:2010ne}.  This inspired the conjecture that the theory enjoys unbroken center symmetry for all $L$. 

If one believes this conjecture, then adjoint QCD enjoys large $N$ volume independence:  roughly speaking, the $L$-dependence of observables vanishes\cite{Kovtun:2007py}, see also \cite{Eguchi:1982nm,Bhanot:1982sh}.    But the IR dynamics of adjoint QCD are strongly coupled on $\mathbb{R}^4$, and volume independence implies that this remains true for any $L$ so long as $L \Lambda \sim N^0$.  So if generic large-distance observables are strongly coupled on $\mathbb{R}^4$, they remain strongly-coupled for any $L$.    So how can one know that a theory enjoys large $N$ volume independence analytically, without  lattice simulations?  Relatedly, just how small does $L$ have to be to make the conclusion of Ref.~\cite{Kovtun:2007py} reliable?

To see the answer,  let us recall the precise statement of large $N$ volume independence on $\mathbb{R}^3 \times S^1$\cite{Kovtun:2007py} for adjoint QCD.  The claim is that correlation functions of topologically-trivial single-trace operators  are independent of $L$ at large $N$ so long as  center symmetry does not break spontaneously.
\footnote{In fact one must also assume that translation symmetry in $S^1$ does not break spontaneously\cite{Kovtun:2007py}.  This is indeed the case in all known examples of Lorentz-invariant theories compactified to $\mathbb{R}^3 \times S^1$, so we do not further discuss this condition.}  
Most of the physics of interest in a gauge theory is in the sector covered by volume independence.  This is conceptually interesting, but also unfortunate for calculations due to strong coupling issues.  But very fortunately, the observables necessary to check the center symmetry realization conditions vital to volume independence are \emph{not} in the volume-independent sector.  
The protototypical operator charged under center symmetry is the Polyakov loop
\begin{align}
\tr \Omega =\tr \mathcal{P} e^{i \oint_{S^1} A}.
\end{align}
This operator is topologically non-trivial because it winds around the compact direction.  When $L\Lambda \gtrsim 1$,  quantum fluctuations in $\tr \Omega^n$ are large and one must appeal to numerical lattice Monte Carlo simulations to determine $\langle \tr \Omega^n\rangle$; the simulations indicate that center symmetry is not broken\cite{Bringoltz:2009kb,Bringoltz:2011by,Azeyanagi:2010ne}.


However, things are simpler when $L$ is sufficiently small to make quantum fluctuations in $\tr \Omega^n$ small, so that a loop expansion becomes useful.    In such a regime, it becomes meaningful to compute the the Coleman-Weinberg effective potential for the holonomy of the gauge field on $S^1_L$.  This effective potential is often called the Gross-Pisarski-Yaffe (GPY) effective potential\cite{Gross:1980br}, and takes the form\cite{Kovtun:2007py}
\begin{align}
V_{\rm eff}(\Omega) = \frac{2(n_F-1)}{\pi^2L^4} \sum_{n\ge 1} \frac{1}{n^4} |\tr\Omega^n|^2 \,,
\label{eq:GPY_adj}
\end{align}
in $U(N)$ adjoint QCD.   
The minimum of this potential for $n_f>1$ is 
\begin{align}
\Omega = e^{i\alpha} \mathrm{diag}(1, \omega, \cdots, \omega^{N-1}), \omega = e^{2\pi i /N}\,.
\end{align}
where $\alpha$ is arbitrary.\footnote{In an $SU(N)$ theory $\alpha$ would be fixed by requiring $\det \Omega = 1$, and one would replace $|\tr \Omega^n|^2$ by $|\tr \Omega^n|^2 -1$ in the effective potential.} At this minimum the traces of the holonomy vanish,
\begin{align}
\langle \tr \Omega^n \rangle = 0 \,,\; n  \neq 0 \textrm{ mod } N 
\end{align}
and the $\mathbb{Z}_N$ center symmetry of adjoint QCD is not spontaneously broken.\footnote{More precisely, the center symmetry is $\mathbb{Z}_N$ in $SU(N)$ adjoint QCD, while in $U(N) = [SU(N) \times U(1)]/\mathbb{Z}_N$ the center symmetry is extended from $\mathbb{Z}_N$ to $U(1)$.  }  This means that by the standard criterion, adjoint QCD is confining whenever the one-loop computation leading to Eq.~\eqref{eq:GPY_adj} is valid.

Clearly the only chance for the calculation to be valid is to take $L$ small enough that one can appeal to asymptotic freedom.  The question is what is the precise criterion involved.  In the 3D effective theory valid for $L \Lambda \ll 1$, the holonomy  ``Higgses" the gauge group $SU(N)\to U(1)^{N-1}$, but the  $W$-boson mass scale is $1/(NL)$.  This means that an Abelianized 3d effective theory, which is weakly coupled\cite{Unsal:2007jx,Unsal:2008ch},  is a valid description only if $NL\Lambda \ll 1$ .    This may make it tempting to conclude that Eq.~\eqref{eq:GPY_adj} is valid only when $NL\Lambda \ll1$. 

This is not correct.  To appreciate this, we need to discuss what controls the corrections to Eq.~\eqref{eq:GPY_adj}.  First, consider the behavior of loop corrections near the center-breaking extrema $\Omega = \omega^{k} \mathbf{1}$.  The physics here is essentially identical to that of thermal YM theory.  It is well known that naive perturbation theory in thermal YM theory with temperature $T = 1/L$ suffers from infrared (IR) divergences starting at three loops (corresponding to $\lambda^2$ terms in an expansion of the free energy $\mathcal{F}$, since the $\lambda^0$ term in $\mathcal{F}$ is a one-loop term).  
The physical origin of these IR divergences are the $\sim N^2$ zero modes of the theory on the circle.  IR divergences in perturbation theory are of course a signal that the theory is trying to develop effective masses for some modes, and the strength of IR divergences is correlated with the size of these effective masses.  Here the IR divergences are cut off by the appearance of effective masses for electric and magnetic gluons.  The `Debye' electric gluon mass is calculable in resummed perturbation theory, $m_{D} \sim \lambda^{1/2} L^{-1}$, while the magnetic gluon mass $m_{\rm mag} \sim \lambda  L^{-1}$ is determined by the confining dynamics of  three-dimensional YM theory and is not calculable in perturbation theory.  Taking these effective masses into account by appropriate resummations produces non-analyticities in the free energy of the form $L^{-4} \lambda^{3/2}$ and $L^{-4} \lambda^2\log \lambda$, where the 't Hooft coupling is taken at the scale $1/L$.  In the end, however, all corrections to the one-loop free energy are small whenever $ L \Lambda \ll 1$.  So one can trust the one-loop value of the free energy at $\Omega = 1$ whenever $L \Lambda \ll 1$.

What about the loop corrections when $\tfrac{1}{N}|\langle \tr\Omega\rangle| \neq 1$, and in particular near the center-symmetric point in holonomy space?  
The key point is that when $\tfrac{1}{N}|\langle \tr\Omega\rangle| \neq 1$, there are only $\sim N$ zero modes on the circle in perturbation theory, corresponding to the Cartan gluons.  
At the center-symmetric point, the Cartan gluons only develop non-perturbatively small masses, $m_{\rm Cartan} \sim e^{-16\pi^2/\lambda}$\cite{Unsal:2007jx,Unsal:2008ch}, see also Ref.~\cite{Polyakov:1976fu}, while all other modes pick up masses proportional to $\sim 1/(NL)$ and $\lambda^{1/2}/(NL)$.  
Note that all of these mass scales are much smaller than in the thermal case. 
This testifies to the fact that the strength of IR divergences decreases when one moves away from the center-breaking point $\Omega = \omega^{k} \mathbf{1}$, and indeed they are smallest when $\Omega$ takes the center-symmetric value.   
Not coincidentally, this is also the point in holonomy space where volume independence sets in at large $N$, and the physics becomes four-dimensional.  
IR divergences are much weaker in four-dimensional theories compared to three-dimensional theories.  In 4D theories the IR divergences  are cut off by non-perturbatively small effective masses $\sim e^{-8\pi/ (b_0 \lambda)}$ where $b_0= 11/3 - 2n_f/3$,  while in 3D theories the IR masses scale with powers of $\lambda$, as discussed above in the context of thermal YM theory.   
All of this implies that it is meaningful to compare the center-broken and center-symmetric extrema of the potential whenever $L\Lambda \ll 1$.

In thermal YM theory, such considerations justify the famous conclusion of Ref.~\cite{Gross:1980br} that YM theory is in a deconfined phase at high temperature.  In adjoint QCD on a circle with periodic boundary conditions, these considerations  imply that center symmetry is \emph{not} broken when $L\Lambda \ll 1$.  This means that large $N$ volume independence applies to adjoint QCD whenever $L\Lambda \ll 1$.  Lattice calculations\cite{Bringoltz:2009kb,Bringoltz:2011by,Azeyanagi:2010ne} show that center symmetry is also preserved when $L \Lambda \gtrsim 1$.  So the evidence supports the conclusion that adjoint QCD enjoys large $N$ volume independence for all $L$.

\subsection{Derivation of the main claim from large $N$ volume independence}
\label{sec:quick_derivation}

These results, along with known features of large  $N$ volume independence, are actually already enough to give a quick derivation of our main result.  First, we recall that large $N$ volume independence is a statement about toroidal compactifications\cite{Eguchi:1982nm,Kovtun:2007py}.  It has not been studied extensively on manifolds with curvature\footnote{Although see Refs.~\cite{Unsal:2007fb,Basar:2014jua,Shaghoulian:2016gol}.}, so in this subsection we consider compactifying adjoint QCD on $T^3 \times S^1$, and assume that the size of $T^3$ is very large, so that we are effectively in the $\mathbb{R}^3 \times S^1$ limit.

Consider the expectation value of the $(-1)^F$-graded energy density:
 \begin{align}
\langle \mathcal{E} \rangle = \partial_L \log \tilde{Z}
\end{align}
in a 4D gauge theory.  In the $L \to \infty$ limit one expects that
\begin{align}
\langle \mathcal{E} \rangle \sim c \Lambda^4
\label{eq:R4E}
\end{align}
where $\Lambda$ is the strong scale and $c$ is a scheme-dependent constant.\footnote{We are assuming that the theory in question does not flow to a non-trivial IR fixed point. In an IR CFT  on $\mathbb{R}^4$,  any reasonable renormalization scheme choice would lead to $\langle \mathcal{E} \rangle = 0$.}  Eq.~\eqref{eq:R4E} is only defined given a choice of regularization and renormalization scheme, and in the following we assume that the scheme does not break center symmetry.  When $L\Lambda \ll 1$, in theories with a gauge group with rank $\sim N$, one expects
\begin{align}
\langle \mathcal{E} \rangle = a_{\rm typical} L^{-4} +  c \Lambda^4 + \cdots
\label{eq:energy_density_generic}
\end{align}
where $a_{\rm typical} \sim N^2$.  In adjoint QCD on $\mathbb{R}^3 \times S^1$ there cannot be any terms proportional to e.g. $L^{p-4}\Lambda^p$ with $p = 1,2,3$ for reasons explained around Eq.~\eqref{eq:general_structure_Z}, without any assumptions about center symmetry.

In a theory which enjoys large $N$ volume independence for any $L\Lambda$, however, the $L^{-4}$ term must vanish, so one must have 
\begin{align}
a = \mathcal{O}(N^{-2})\,.
\label{eq:coefficient_scaling}
\end{align}
Indeed, following Gross and Kitazawa\cite{Gross:1982at}\footnote{See also the literature on twisted Eguchi-Kawai (EK) reduction\cite{GonzalezArroyo:1982ub,GonzalezArroyo:1982hz,GonzalezArroyo:2010ss,Gonzalez-Arroyo:2013bta,Perez:2013dra,Perez:2014sqa}. }, we deduce that  planar perturbation theory in a center-symmetric holonomy background  with just one compact dimension depends on $L$ only through the parameter $LN$.
 Trading $L$ for $LN$ in  Eq.~\eqref{eq:energy_density_generic} with $a_{\rm typical} \sim N^2$ one lands on an effective value of $a$ as given in Eq.~\eqref{eq:coefficient_scaling}.

To recover  Eq.~\eqref{eq:SUSYscaling} we just integrate Eq.~\eqref{eq:R4E} with respect to $L$.  This produces Eq.~\eqref{eq:SUSYscaling} with $\ell = 0$ because we have taken the spatial manifold to be flat for this discussion.    We expect that a careful study of large $N$ volume independence on product manifolds where one of the factors is curved, such as $S^3_R\times S^1_L$, will show that $L$ continues to enter observables in the combination $LN$.  The curvature  term in the effective action will then produce an $N$-independent  term scaling as $1/L^2$ in Eq.~\eqref{eq:energy_density_generic}, and hence reproduce  Eq.~\eqref{eq:SUSYscaling}  with $\ell \sim R$.  This expectation is supported by an explicit large $N$ calculation of $\log \tilde{Z}$ on $S^3 \times S^1$ with small $S^3$ radius $R$.  This calculation is presented in Appendix~\ref{sec:appendix}.

One may ask how the disappearance of the $1/L^3$ term in $\log \tilde{Z}$ can be consistent with the microscopic counting of the degrees of freedom in adjoint QCD.  The answer can be seen from Eq.~\eqref{eq:GPY_adj}.  A center-symmetric holonomy means that different color components of the gluons and adjoint quarks do not contribute equally to $\log \tilde{Z}$.  Indeed, let $e^{i\phi_{a}}, a = 1, \ldots, N$ be the eigenvalues of $\Omega$.  Then one can write the one-loop effective potential as
\begin{align}
V_{\rm eff}(\Omega) &= \frac{2 (n_F -1)}{L} \sum_{a,b=1}^{N}\int \frac{d^3p}{(2\pi)^2} \log
\left(1-e^{-p L} e^{i \phi_a - i\phi_b}\right) 
\label{eq:colorsum}\\
&=  \frac{2(n_F-1)}{\pi^2L^4 }  \sum_{n = 1}^{\infty} \sum_{a,b=1}^{N} \frac{1}{n^4} e^{i n(\phi_a  - \phi_b )} 
 = \frac{2(n_F-1)}{\pi^2L^4 } \sum_{n=1}^{\infty} \frac{1}{n^4} |\tr \Omega^n|^2  \,. \nonumber
\end{align}
The sum over $a, b$ in Eq.~\eqref{eq:colorsum} is a sum over the color for the gluons and adjoint quarks.  A non-trivial holonomy can be interpreted as giving twisted boundary conditions for these fields, and consequently their contributions to the partition function come with phases determined by the holonomy.  
Evaluating this expression on its center-symmetric minimum gives $\log \tilde{Z}$, and the quark and gluon contributions get weighed by phases which are $N$-th roots of unity.  This causes very strong destructive interference, and leads to the one-loop result
\begin{align}
\log \tilde{Z} &= \frac{2(n_F-1)}{\pi^2L^4 } \sum_{k = 1} \frac{1}{N^4 k^4} N^2  = \frac{2(n_F-1)}{\pi^2L^4 N^2} \zeta(4)  \,,
\label{eq:potential_punchline}
\end{align}
rather than $V_{\rm eff} \sim (n_f -1) N^2 L^{-4}$, which would have held if center symmetry were broken.  In the Section~\ref{sec:cancellations} we generalize this one-loop argument to all orders in the perturbative expansion, and explain why non-perturbative effects cannot change the results.

\section{Holonomy effective potential and cancellations}
\label{sec:cancellations}
In this section we complement the arguments of Section~\ref{sec:quick_derivation} by a more explicit discussion of the small $L$ behavior of the $\log \tilde{Z}$ in adjoint QCD. 

\subsection{General structure}
To understand the structure of $\tilde{Z}$ for small $L$, one can integrate out all modes with energies $\gtrsim 1/L$. To avoid IR divergences, we put the theory on a compact spatial manifold $M$.  In adjoint QCD, $\tilde{Z}$ must take the form\cite{DiPietro:2014bca}
\begin{align}
\log \tilde{Z} &= a(N,\lambda) L^{-3} \int d^{3}x \sqrt{g} + b(N,\lambda) L^{-1} \int d^{3}x \sqrt{g} \, \mathcal{R} \\
&+ c(N,\lambda) L \int d^{3}x \sqrt{g}  \, \mathcal{L}_{\rm eff} + \cdots\nonumber
\label{eq:general_structure_Z}
\end{align}
Here $a, b$ and $c$ are functions determined by matching to the UV theory,  $g$ is the metric on $M$, $\mathcal{R}$ is the Ricci scalar curvature associated to $g$, and $\mathcal{L}_{\rm eff}$ is the effective Lagrangian for modes which are massless on the scale $L$.  
In adjoint QCD
 \begin{align}
 \mathcal{L}_{\rm 3d} = \frac{1}{2g^2} \left[\tr F_{i j}F^{i j} + \sum_{a = 1}^{n_f} \bar{\lambda}_a \Dslash_{\rm spatial} \lambda_{a}\right] \,,
 \end{align}
  where $i,j$ are 3d indices.  Note that gauge-invariance forbids terms like $L^{-1} \int d^3{x} \sqrt{g}\, \mathcal{O}_2$ where $\mathcal{O}_2$ is a dimension-$2$ local operator built out of gluons and quarks.  There cannot be any term like  $L^{-2} \int d^3{x} \sqrt{g}\, \mathcal{O}_1$ with an operator built out of dynamical or background fields with dimension-$1$, because there are no such operators consistent with the symmetries.  Finally, the term $L^{0} \int d^3{x} \sqrt{g} \,\mathcal{O}_3$ is also forbidden, because the only candidate dimension $3$ operator ${\lambda}\lambda$, where $\lambda$ is an adjoint Weyl fermion in 4d,   transforms  under chiral symmetry.    

The coefficients of the volume term $L^{-3}$ and the curvature term $L^{-1}$ are scheme-independent and their values are physical.  In thermal YM theory, for example, $a(N,\lambda)$ is just the coefficient of $T^{4}$ in the high-temperature expansion of the free energy density $\mathcal{F}$
 \begin{align}
 \mathcal{F}_{\rm YM} = \frac{1}{\beta V} \log Z = a(N,\lambda)T^4 ,
 \end{align}
 and $a \sim N^2$.
 
 
Here we are dealing with a theory which remains confining for small $L$. A naive microscopic count of the degrees of freedom would lead one to expect
\begin{align}
a(N,\lambda) \overset{!}{=} \mathcal{O}(N^2) \,.
\end{align}   
But in the confining phase it is expected that the free energy scales as $N^0$. In a theory where confinement persists to small $L\Lambda$, this already means that the growth of $a$ with $N$ cannot be stronger than
\begin{align}
a(N,\lambda) = \mathcal{O}(N^0) \,.
\end{align}
This already requires some highly non-trivial cancellations. 
Our goal here is, of course, to argue that in adjoint QCD the cancellation are even stronger, and in fact
\begin{align}
a(N,\lambda) = \mathcal{O}\left(\frac{1}{N^{2}}\right) \,.
\label{eq:goal}
\end{align} 

\subsection{Holonomy effective potential to all orders}
Rather than constraining $a(N,\lambda)$ directly, we will instead discuss the structure of the effective potential for the holonomy, $V_{\rm eff}(\Omega)$:
 \begin{align}
 V_{\rm eff}(\Omega) = a(N,\lambda,\Omega) L^{-4} + \cdots
 \label{eq:vScalingSetup}
 \end{align}
where $\cdots$ are finite-volume corrections and corrections involving positive powers of the strong scale $\Lambda$.  
When $V_{\rm eff}(\Omega)$  is minimized with respect to $\Omega$, it coincides with the $(-1)^F$-graded free energy density, and so a study of $V_{\rm eff}(\Omega)$ gives us information about the  function we are really after, namely $a(N,\lambda)$.  
We find it easier to understand the implications of center symmetry starting with $V_{\rm eff}$ rather than  working with $a(N,\lambda)$ directly.

We now record some important basic observations concerning the structure and physical origin of $a(N, \lambda, \Omega)$.  First, $a(N, \lambda, \Omega)$ is fully determined in perturbation theory. Dimensional transmutation means that non-perturbative effects, which are weighed by positive powers of $e^{-1/\lambda}$, generate contributions involving positive powers of the strong scale $\Lambda$.  This means that non-perturbative effects cannot contribute to the coefficient of $1/L^4$ in $V_{\rm eff}$, so we only need to consider the perturbative effects from here onward.

Second, gauge invariance along with the definition of quantum effective potentials implies that the dependence of $f$ on $\Omega$ can only be through the variables 
\begin{align}
u_n \equiv \tfrac{1}{N}  \langle \tr \Omega^n \rangle,  \; n \in \mathbb{Z} \,.
\end{align}
We have normalized these variables so that  $u_n \sim \mathcal{O}(1)$ at large $N$.     Next, standard large $N$ arguments imply that $a$ has an expansion\footnote{This expansion in $1/N^2$ is expected to be asymptotic, see e.g.~\cite{Shenker:1990uf}, but this does not matter for our argument.} in inverse powers of $N^2$:
\begin{align}
a(N, \lambda, \{u_n\})=  N^2 a_{0}(\lambda,\{u_n\})+N^0 a_{1}(\lambda,\{u_n\})+\mathcal{O}(N^{-2})
\label{eq:extensiveV}
\end{align}
where the functions $a_{g}$ are sums of Feynman diagrams of genus $g$.  The functions $a_{g}$ become manifestly $N$-independent when $|u_n| = 1$ and center symmetry is broken.  We will see that when the holonomy deviates away from the center-broken locus, 
\begin{align}
|u_{n}| =1 \,, \; \ \forall \, n \in \mathbb{N} \,,
\end{align}
 the functions $a_g$ decrease in magnitude.  In particular, when center symmetry is unbroken and  $u_{n} = 0$ for all $n \neq 0\,\, \textrm{mod}\,\, N$, we will see that $a_g$ become so small that $a(N,\lambda)$ goes to zero at large $N$.  

To see how this comes about, consider the expansion of the functions $a_g$ as formal power series in $\lambda$\footnote{%
This expansion in $\lambda$ is also asymptotic due to e.g. IR renormalon effects, but this is also irrelevant to our argument, because renormalons only matter for understanding the terms involving powers of $\Lambda$. }
\begin{align}
a_{g} (\lambda,\{u_n\})= \sum_{p = 0}^{\infty} \lambda^{p} c_{g,p}(\{u_n\}).
\end{align}
 The coefficients $c_{g,p}$ are functions of the holonomies.  Feynman diagrams involving $p$ powers of the 't Hooft coupling at genus $g$ have $p+2-g$ index loops, and to contribute an $L$-dependent piece to $V_{\rm eff}$ (and hence to $\log \tilde{Z}$), at least one of the propagators in the position-space representation of the diagram has to go around the circle.   So the relevant diagrams at order $\lambda^{p}$ produce expressions involving at least $2$ and at most $p+2-g$ holonomy traces. Finally, center symmetry implies that the sum of the powers of these holonomies must add up to zero.  All this taken together means that we can write
\begin{align}
c_{g,\,p}(\{u_n\}) &=  \sum_{n \in \mathbb{Z}_{+}}
                  c^{g,p}_{2}(\vec{n}) u_{n}u_{-n}\\
                  &+\sum_{\substack{
 \vec{n} \in \mathbb{Z}^{2}  \\
                  n_1 \neq 0, \, n_2 \neq 0
                  }}
                  c^{g,p}_{3}(\vec{n}) u_{n_1}u_{n_2} u_{-n_1-n_2} \nonumber \\
&+ \ldots \nonumber \\              
&+ \sum_{
 \substack{
 \vec{n} \in \mathbb{Z}^{p+1 - g}  \\
                  n_{i} \neq 0
                  }
         } 
                  c^{g,p}_{p+2-g}(\vec{n}) u_{n_1}u_{n_2} \cdots u_{n_{p+1 - g }} u_{-n_1-\ldots - n_{p+1 - g}} \,,\label{eq:holonomy_expansion}
\end{align}
where we have separated terms with different numbers of holonomy traces.  The dependence of $c^{g,p}_{p+2-g}(\vec{n})$ on $\vec{n}$ is constrained by noting that  \begin{align}
 c^{g,p}_k(n_1, n_2, \ldots, n_{k}) = c^{g,p}(n_{\mathcal{P}(1)}, n_{\mathcal{P}(2)}, \ldots, n_{\mathcal{P}(k)}) \,, 
 \end{align}
 where $\mathcal{P}$ is an arbitrary permutation, since
 \begin{align}
 u_{n_1}u_{n_2} \cdots u_{n_{k}} u_{-n_1-\ldots - n_{p+1 - g}}  = u_{n_{\mathcal{P}(1)}}u_{n_{\mathcal{P}(2)}} \cdots u_{n_{\mathcal{P}(p+1 - g )}} u_{-n_{\mathcal{P}(1)}-\ldots - n_{\mathcal{P}(k)}}    \, .
 \end{align}
 This means that the summands entering our ansatz are effectively `spherically symmetric'.

 Our goal is to establish bounds on the $\vec{n}$-dependence of the functions $c^{g,p}$ which have the effect of ensuring Eq.~\eqref{eq:goal}.   To find such bounds, we observe  the effective potential must make sense for all values of $u_n$, including $|u_n| =1$.  This can only work if the coefficients $c^{g,p}_{k}(\vec{n})$ have sufficiently fast fall-off at large $|n|$. Requiring
 \begin{align}
\sum_{\substack{
 \vec{n} \in \mathbb{Z}^{k-1} }}c(\vec{n})
 \end{align}
 to converge then places restrictions on the large-$|n|$ scaling of the summand $c$.  The permutation-symmetry property means that conditionally-convergent expressions such as
  \begin{align}
 \sum_{\vec{n} \in\mathbb{Z}^2 } \frac{n_1^2 - n_2^2}{n_1^4+n_2^4}
 \end{align}
 cannot appear.   So the sums we are dealing with must converge absolutely.  The  expressions with $k$ holonomy traces are $k-1$ dimensional sums, and convergence requires the associated coefficient functions to scale as 
 \begin{align}
 c^{g,p}_{k}(\vec{n}) \sim \frac{1}{|n|^c}, \; c>k-1.
 \label{eq:convergence}
 \end{align}
This is as far as we have been able to get in deriving general constraints on $V_{\rm eff}$.  

However, for expressions involving only two holonomy traces
 \begin{align}
 \sum_{n \in \mathbb{Z}_{+}}
                  c^{g,p}_{2}(n) u_{n}u_{-n}
                  \label{eq:two_traces}
 \end{align}
  one can show a stronger constraint, which will be important below.\footnote{We are very grateful to L. G. Yaffe for suggesting the argument which follows.}    Consider $\partial_L V_{\rm eff}(\Omega)$. When evaluated  on the minimum of $\Omega$, this computes the expectation value of the energy density.   In perturbation theory, a holonomy dependence involving two traces arises when precisely one gluon or adjoint quark propagator goes around the circle $S^1$, while the rest do not.  If we denote the position-space gluon propagator on $\mathbb{R}^4$ by $G(x^{\mu})^{ab}$, where all labels except color have been suppressed, then the $\mathbb{R}^3 \times S^1$ propagator can be written as
\begin{align}
G(x^{\mu}; L; \Omega)^{ab} = \sum_{n \in \mathbb{Z}} G(x^{\mu}+ n L \delta_{4,\mu})^{ab} e^{i( \alpha_a - \alpha_b) n}
\end{align} 
where $\Omega \sim \mathrm{diag}(e^{i \alpha_1}, \ldots,e^{i \alpha_1})$.  This is just a sum-over-images construction of a periodic function from a non-periodic one.   For the present application, where we are interested in the finite-volume contribution to the partition function, we need to consider propagators where $x^{\mu}$ vanishes.  This leads to UV divergences, but they are the same as on $\mathbb{R}^4$ and can be ignored, since we are really after $\tilde{Z}(L)/\tilde{Z}(L \to \infty)$.  Passing to momentum space in $\mathbb{R}^3$, we are led to consider expressions of the form
\begin{align}
c^{g,p}_{2}  \sim \int d^{3}p \sum^{N}_{a,b = 1} f 
[G^{a c}(p; L; \Omega)] g(\vec{p})_{c b}
\end{align}
and $g(\vec{p})$ encapsulates the contributions of loops of gluons and quarks whose propagators do not go around $S^1$, and $f$ is some linear function acting on the $S^1$ gluon propagator, which can involve derivatives.   Since only one gluon propagator goes around $S^1$, this expression involves two traces of the color holonomy.  Moreover, the circle size $L$ always enters the expressions together with $n$ in the combination $nL$.  We are interested in the terms which scale as $1/L^4$, and putting these observations together we learn that Feynman diagrams where a single gluon goes around $S^1$ always produce contributions that scale as $1/n^4$. \footnote{Indeed, this was seen by explicit calculation of two-loop contributions to the effective potential in \cite{Dumitru:2013xna}.}  This means that the function $f$ must involve two derivatives.  This is quite natural, seeing as the lowest-dimension gauge-invariant and Lorentz-invariant operator in YM theory, $\tr F^2$, has dimension $4$.     It is not hard to check that  the discussion above also applies if we replace $G$ with an adjoint quark propagator.    So we conclude that 
\begin{align}
c^{g,p}_{2}(n) \sim \frac{1}{n^4} \,.
\end{align}
Note that this is much better than the $1/n^{1+\delta}, \delta>0 $ scaling required for convergence.

\subsection{Cancellations due to center symmetry}

Now we are finally in a position to collect some rewards from the long  discussion above.  We already know from Section~\ref{sec:volume_independence} that the holonomy takes a center-symmetric expectation value in adjoint QCD, so that 
\begin{align}
|u_n| = 
\begin{cases}
	0  &n \neq 0 \textrm{  mod  } N \\
	1 & n = 0 \textrm{  mod  } N
\end{cases}
\label{eq:center_symmetry}
\end{align}
In Section~\ref{sec:quick_derivation} we saw that this leads to a $1/N^4$ suppression in the one-loop effective potential relative to its naive $N^2$ scaling.   Using the result at the end of the preceding subsection, exactly the same suppression appears for all terms involving two holonomy traces in planar perturbation theory.  Terms with two traces from genus-one diagrams are (of course) even more suppressed.  

What about terms with more than two traces of the holonomy?  The convergence constraint in Eq.~ Eq.~\eqref{eq:convergence}  implies that these terms, which are multiplied by $N^2$, must go to zero \emph{faster} than $N^2$ when Eq.~\eqref{eq:center_symmetry} holds.  Since the $1/N$ expansion is organized in powers of $1/N^2$, this means that all of these terms must go to zero at least as fast as $1/N^2$.  The same remarks apply to the genus one and higher diagrams.  So  in adjoint QCD all perturbative contributions to the $1/L^3$ coefficient in $\log \tilde{Z}$ vanish as $N \to \infty$.

Non-perturbative effects cannot contribute to the coefficient of $1/L^3$ in $\log \tilde{Z}$.  So we conclude that the coefficient of the extensive $VL^{-3}$ term in the small-$L$ expansion of $\log \tilde{Z}$ vanishes as $1/N^2$ for any $n_f>1$ in the large $N$ limit.   This matches the general expectations from large $N$ volume independence explained in Section~\ref{sec:quick_derivation}.  Of course, the coefficient of $1/L^3$ also vanishes when $n_f=1$ at any $N$, due to supersymmetry.    So,  at least for the specific observable we have been discussing, the large $N$ spectral conspiracy in adjoint QCD is just as powerful as supersymmetry!

\section{Deformations}
\label{sec:deformations}
We have now seen that massless $U(N)$ adjoint QCD features remarkably powerful cancellations between its bosonic and fermionic excitations, leading to Eq.~\eqref{eq:UNresult} when one computes a $(-1)^F$-graded partition function.  In this section we discuss what happens if we introduce gradings by symmetries other than fermion number, or take the quark masses away from zero.  

\subsection{Alternative gradings and comments on chiral phase transitions}
Let us see how the phase structure of adjoint QCD depends on gradings by global symmetries other than $(-1)^F$, especially grading by the flavor symmetries.  Along the way we will also explain the conditions for a smooth dependence of the theory on the compactification scale $L$ regardless of the realization of  chiral symmetry on $\mathbb{R}^4$.  

For simplicity, let us first consider massless adjoint QCD with $n_f=2$.  There is an $SU(2)$ continuous chiral symmetry\footnote{
 There are  subtleties  when $N=2$, see \cite{Cordova:2018acb}  for a careful discussion.} and a discrete $\mathbb{Z}_{4N}$ chiral symmetry, as well as a $\mathbb{Z}_N$ one-form center symmetry.  Thanks to a mixed 't Hooft anomaly, the $\mathbb{Z}_{4N}$ axial symmetry is spontaneously broken whenever center symmetry is  unbroken \cite{Komargodski:2017smk,Shimizu:2017asf}.     On $\mathbb{R}^4$, the continuous chiral symmetry might be spontaneously broken to the maximal vector-like subgroup, $SO(2)$.  (This is the widely-held expectation for generic values of $N$.)

Let $\psi$ be a flavor doublet,
\begin{align}
\psi = \left( \begin{array}{c} \psi_1\\
\psi_2 \end{array}  \right).
\label{eq:doublet}
\end{align}
When compactifying the theory, one can consider flavor-twisted boundary conditions $\psi(x_3+L) = g \psi(x_3)$, where $g \in SO(2)$.  The matrix $g$ can be diagonalized without loss of generality by using flavor rotations, giving a one-parameter family of boundary conditions\footnote{The same setup was explored  in Ref.~\cite{Misumi:2014raa}, where the holonomy effective potential was also written down and analyzed numerically. The emphasis of our analysis is different, but  it agrees with Ref.~\cite{Misumi:2014raa}  in  areas of overlap. }
\begin{align}
\psi(x_3+L) = 
\begin{pmatrix}
e^{i\varphi} & 0 \\
0 & e^{-i \varphi}
\end{pmatrix}
\psi(x_3) \,.
\label{eq:twisted_BC}
\end{align}

Note that $\varphi = \pi$ corresponds anti-periodic `thermal' boundary conditions, while $\varphi = 0$ corresponds to periodic `spatial' boundary conditions.  Turning on a generic twist angle $\varphi$ is equivalent to working with periodic quark fields  with a background flavor $SO(2)$ holonomy
\begin{align}
\begin{pmatrix}
e^{i\varphi} & 0 \\
0 & e^{-i \varphi}
\end{pmatrix}= U = \mathcal{P} e^{i \oint_{S^1} \mathcal{A}_4} \,,
\end{align}
where $\mathcal{A}$ is the background flavor gauge field. This is also equivalent to turning on an imaginary chemical potential $i\varphi/L$ for the charge associated to the Cartan subgroup $U(1) \subset SU(2)$.  Physically, one can package the two Weyl fermion flavors $\psi_1 $ and $\psi_2$ into a Dirac fermion $\Psi$.
Then the fermion number symmetry 
 $U(1)_F$ is isomorphic to the   vector-like $SO(2)= U(1) $ subgroup of $SU(2)$, which remains unbroken. Hence,
 \begin{equation}
 F = {Q}_{U(1)}\,,
 \end{equation}
 where $F$ is the fermion charge.   The Euclidean path integral with the boundary condition \eqref{eq:twisted_BC} computes a twisted partition function
\begin{equation}
\tilde{Z}(L, \varphi) = \tr (-1)^F e^{-L \hat{H}} e^{i\varphi \hat{Q}_{U(1)_F}}
\label{35}
\end{equation}
where $\hat{H}$ is the Hamiltonian operator while $\hat{Q}_{U(1)}$ is the charge operator for the $U(1)$ symmetry.

Suppose that the quarks are massless.  Then the boundary condition \eqref{eq:twisted_BC} explicitly breaks the flavor symmetry from $SU(2)_F$ to $U(1)_F \equiv SO(2)$ for any finite $L$ and $\varphi \neq 0, \pi$.  If continuous chiral symmetry is spontaneously broken on $\mathbb{R}^4$, then the breaking pattern must be $SU(2) \to SO(2)$, so that on $\mathbb{R}^4$ one would get two exactly massless ``diquark"  Nambu-Goldstone bosons, $G^{\pm}$.     
Given our choice of boundary conditions, at finite $L$ these Nambu-Goldstone bosons  pick up effective masses $\sim \varphi/L$. So with \eqref{eq:twisted_BC} we  should \emph{not} expect to see any exactly gapless bosons in the spectrum of the theory on $\mathbb{R}^3 \times S^1$, because there is no exact continuous symmetry which could be spontaneously broken.\footnote{$SO(2)$ is vector-like symmetry and the present theory has a non-negative path integral measure\cite{Vafa:1983tf}, so $SO(2)$ cannot break spontaneously.}  Consequently, the realization of the unbroken continuous flavor symmetry must be identical for all $L$ with the boundary condition of Eq.~\eqref{eq:twisted_BC} so long as $\varphi \neq 0,\pi$.

At large $L\Lambda$, we expect an unbroken center symmetry, so that 
\begin{align}
\langle \tr \Omega^n  \rangle = 0,\; n = 1,\ldots, N-1.
\end{align}
where $\Omega$ is the Polyakov loop around $S^1$.  Let us now examine the realization of center symmetry at small $L\Lambda$.  Generalizing the GPY calculation of the holonomy effective potential\cite{Gross:1980br}, we find the one-loop effective potential for $\Omega$ with massless quarks: 
\begin{align}
V_{\rm eff}(\Omega; \varphi) = 
\frac{2}{\pi^2L^4} \sum_{n\ge 1} \left[-1+  2\cos(n\varphi) \right] \frac{ |\tr\Omega^n|^2 }{n^4}\,.
\label{master} 
\end{align}
The first term  comes from the gluons while the second term comes from the adjoint fermions.  

To determine the phase of the theory, one only has to specify the first $\lfloor N/2 \rfloor$ expectation values powers of the holonomy, since this suffices for the determination of the $\Z_N$ center symmetry. 
  Thus, to preserve center symmetry, we have to make sure that the masses of the Wilson lines  for 
 $\tr \Omega^k $ are positive for $k < \lfloor N/2 \rfloor$.  This gives the condition 
\begin{align}
- &1+  2\cos(k\varphi) >0, \qquad  k=1,2, \ldots, \lfloor N/2 \rfloor
\label{eq:alpha_condition}
\end{align}
for center symmetry to be preserved at small $L\Lambda$. Consequently, so long as the twist $\varphi$ obeys the condition
\begin{align}
0 < |\varphi| < \frac{2\pi}{3N} \,,
\label{eq:twist_condition}
\end{align}
then the large $L$ and small $L$ regimes of the theory are smoothly connected, in the sense that all order parameters for all of the symmetries  --- center symmetry and the discrete and continuous chiral symmetries --- are realized in the same way for any $L$.  However, in the large $N$ limit with $L\Lambda$ fixed, the only boundary condition/partition function grading which preserves confinement at small $L\Lambda$ is $\varphi = 0$.  Grading by anything other than fermion number destroys the cancellations described in the preceding sections.

While we set $n_f=2$ above,  the construction easily generalizes to higher $n_f$, where  the choice of boundary conditions can be parameterized by $n_f-1$ angles.  The one-loop effective potential becomes
 \begin{align}
V_{\rm eff}(\Omega; \varphi_i) &= 
\frac{2}{\pi^2L^4} \sum_{n\ge 1} \left[-1+   \cos(n\varphi_1) +\ldots + \cos(n\varphi_{n_f-1}) +\cos(n( \varphi_1+\cdots+ \varphi_{n_f-1}))  \right]  \nonumber\\
&\times \frac{|\tr\Omega^n|^2-1}{n^4}\,.
\label{eq:genericNf} 
\end{align}
At finite $N$, one can always find non-coincident angles $\varphi_{i}$ which preserve center symmetry.  At large $N$, it is much harder. To see why, first note that the fermions produce the largest repulsion for eigenvalues of $\Omega$ when all of the angles are set to zero.  We have already analyzed this case in the preceding sections.  The next largest repulsion can be obtained when all but one angle, say $\varphi_{1}$, are set to $0$. In this case
\begin{align}
V_{\rm eff}(\Omega; \varphi_1) = 
\frac{2}{\pi^2L^4} \sum_{n\ge 1} \left[(N_F-3) + 2 \cos(n\varphi_1) \right]
&\times \frac{|\tr\Omega^n|^2-1}{n^4}\,. 
\end{align}
If $n_f \le 4$ and we pick $\varphi_1 \sim \mathcal{O}(1)$, then there will exist an $n$ such that $n_f-3 + 2\cos(n\varphi_1) < 0$.  This means that at large $N$, center symmetry breaks. The only way to avoid this is to take $\varphi_1 \sim 1/N$, but then at large $N$ one again lands on periodic boundary conditions/$(-1)^F$ grading as the only way to ensure confinement at small $L\Lambda$.

The case of $n_f = 5$ is special for several reasons. For example, it is widely believed that when $n_f = 5$, adjoint QCD is in the conformal window on $\mathbb{R}^4$.  The more important point for the present discussion is that  $n_f - 3 + 2\cos(n\varphi_1) = 2(1+\cos(n\varphi_1)) \ge 0$ when $n_f =5$, and for large $N$, when large values of $n$ become important for  center symmetry realization, $2(1+\cos(n\varphi_1))$ can get arbitrarily close to $0$.  This means that  the one-loop potential can become very small.  So the fate of center symmetry in $n_f =5$ adjoint QCD with additional gradings on top of $(-1)^F$ is sensitive to higher loop corrections, and is left to future work.   We do not consider  $n_f >5$, because then the theory is not asymptotically free and the small-circle limit is strongly coupled.

\subsection{Mass deformations}
So far we have kept the adjoint quarks massless.  What happens to center symmetry at small $L$ if we lift this assumption? The holonomy effective potential takes the form\cite{Unsal:2010qh}
\begin{align}
V_{\rm eff}(\Omega) &= \frac{2}{\pi^2L^4} \sum_{n\ge 1}\left(-1
+ \frac{1}{2}\sum_{a = 1}^{N_F} (n L m_a)^2 K_2(n L m_a) \right)  \frac{|\tr\Omega^n|^2 }{n^4} 
\label{eq:GPY_massive}
\end{align}
The $-1$ term is generated by the gluons, while the term involving Bessel functions $K_2$ is generated by the adjoint fermions.   

If all of the quarks have a common mass $m$, the fermionic contribution to the potential for the large holonomy windings $n \sim N$ is exponentially suppressed thanks to $K_2(N L m) \sim e^{-N L m}$, and the gluons force center symmetry breaking.  Then the coefficient of $L^{-3}$ in $\log \tilde{Z}$ scales as $N^2$.  The only way to avoid this is to take $m \sim 1/N$, but at large $N$ this is the same as setting $m=0$.

If $N_F = 1$ and $m_1>0$, the effective potential is minimized for $\frac{1}{N} |\langle \tr \Omega\rangle| = 1 $, corresponding to spontaneously broken center symmetry.  The only way to protect center symmetry at small $L$ in the one-flavor theory is to take $m_1 = 0$.   This amounts to going to the supersymmetric $\mathcal{N} = 1$ SYM theory, and there it is known that\cite{Davies:1999uw}
\begin{align}
V_{\rm eff}(\Omega) = 0,\;\; \textrm{perturbation theory}
\end{align}
The minimum of the full non-perturbative potential has the property that $\tr\Omega^n = 0$ for all $n \neq N k$\cite{Davies:1999uw,Poppitz:2012sw}.   So for $m_q = 0$ the coefficient of $L^{-3}$ in $\log \tilde{Z}$ vanishes trivially due to supersymmetry, but if $m_q >0$ it does not vanish and scales as $N^2$.

Now let us suppose that $n_f >1$,  one of the flavors of quarks is massless, $m_1 = 0$, but all of the other quarks have non-vanishing masses.   Suppose for simplicity that all of the non-vanishing quark masses have a common mass $m$.  Then  Eq.~\eqref{eq:GPY_massive} shows that the one-loop gluon contribution is cancelled by the contribution of the massless quark flavor, while the remaining quarks make positive-definite contributions to $V_{\rm eff}(\Omega)$.  It is then tempting to say that center symmetry is stabilized.  This will indeed be self-consistently true at finite $N$ within the domain of validity of the one-loop calculation so long as $m L \ll 1$.

But life is harder at large $N$, because we must stabilize $\sim N$ winding modes of the holonomy.  When $mL \ll 1$,  one can be sure that $\tr \Omega^n$ with $n \sim \mathcal{O}(1)$ will experience a center-stabilizing potential.  But it is not clear what happens to $\tr \Omega^n$ with $n \sim \mathcal{O}(N)$, because for such modes the one-loop effective potential vanishes exponentially in $N$.  The non-perturbative neutral bion center-stabilization mechanism of $\mathcal{N}=1$ super-YM theory is only under control when $NL\Lambda \ll 1$, so we cannot appeal to it.  Moreover, without SUSY, we have no way to argue that all perturbative contributions to the holonomy effective potential cancel, nor do we know how to control their overall sign for the high-winding modes.  It thus appears that the fate of center symmetry in adjoint QCD with precisely one massless quark flavor rests on the explicit evaluation of higher-loop contributions to the effective potential.  There are two possibilities:  either these contributions favor center symmetry breaking, which would imply that at small $L\Lambda$ the theory is in a `partially-confined' phase, or it is exactly confining with an unbroken center symmetry.  In the latter case the cancellations we've observed in the massless theory would hold, while in the former case they would not.  

Finally, suppose that $n_f > 2$, and two (or more) quark flavors have vanishing masses, while the rest do not.  One can then see that all windings of the holonomy have $\mathcal{O}(1)$ positive effective masses for any $N$. As a result, center symmetry will be preserved for $L\Lambda \ll 1$, and the cancellations we saw in the fully massless theory will continue to hold.  But of course, such theories interpolate between multi-flavor adjoint QCD with different numbers of massless flavors, so this result is very natural.

All of this suggests that the class of non-supersymmetric theories obeying Eq.~\eqref{eq:SUSYscaling} is larger than just massless adjoint QCD.  It would be very interesting to understand which theories should obey  Eq.~\eqref{eq:SUSYscaling}  more systematically.

\section{Discussion}
\label{sec:discussion}
In the preceding sections we have established that adjoint QCD features extremely precise cancellations in its $(-1)^F$ graded partition function.  We gave two arguments for this result:  a general argument from large $N$ volume independence and a more concrete argument from the structure of the perturbative expansion of the holonomy effective potential.   In this section we discuss some implications of these results. 

First, we discuss the interpretation of the cancellations from the perspective of the hadronic color-singlet excitations of the theory, making a connection with Ref.~\cite{Basar:2013sza}, and highlight why such cancellations are much more difficult to arrange than one might guess.  We then draw a parallel between our  field-theoretic findings and some some properties of non-supersymmetric string theories discussed in Refs.~\cite{Kutasov:1990sv,Dienes:1994jt,Dienes:1994np,Dienes:1995pm}.   

Next, we turn to applications.   First, we discuss why one might hope to derive some implications of our results on large $N$ adjoint QCD to real QCD with fundamental fermions and $N=3$.  Second, we discuss the vacuum energy of adjoint QCD.
A famous implication of supersymmetry is that the vacuum energy vanishes unless supersymmetry is spontaneously broken.  It turns out that something similar takes place in adjoint QCD in the large $N$ limit.  

\subsection{Large $N$ spectral conspiracy and a 4d-2d relation}
\label{sec:spectral_conspiracy}

The key feature of adjoint QCD which is used in our work is that it enjoys large $N$ volume independence when compactified on a circle with periodic boundary conditions.   A corollary is that  the dependence on the circle size is smooth.  It is interesting to understand the implications of this weaker statement.   Recall that the $(-1)^F$-graded partition function can be written as
\begin{align}
\tilde{Z}(L) = \int dE \left[\rho_B(E) - \rho_F(E) \right] e^{-L E}
\end{align}  
Confining large $N$ gauge theories are expected to have densities of states with Hagedorn scaling:  an exponentially-growing density of bosonic states $\rho_{B} \to e^{+L_{H,B} E}$ for large $E$.  Gauge theories with adjoint fermions have light fermionic states, and consequently one also expects $\rho_{F} \to e^{+L_{H,F} E}$, with $L_{H,B}, L_{H,F} \sim N^0$.  If the difference between the bosonic and fermionic densities of states scales exponentially with energy, $\rho_{B} - \rho_F \to e^{+L'_{H} E}$ for some $L'_H \sim N^0$, then $\tilde Z(L)$ \emph{must} have a singularity at some $L_* \le L'_H$.  So unless all exponential growth cancels between the bosons and fermions, smoothness of the physics as a function of $L$ for all $L$ is \emph{impossible} at large $N$. 

Cancellation of all exponential growth is extremely difficult to achieve.  In a large $N$ gauge theory, one can write densities of states as transseries\footnote{Strictly speaking, the large $N$ density of states is not a smooth function.  In the infinite volume limit, it has step-function-type discontinuities associated with thresholds for accessing new hadronic states, which are all stable at large $N$.    When expanded in $1/E$, these step function discontinuities map to oscillatory terms in the expansion, with an oscillation frequency $\sim \Lambda$ in confining theories. In writing the Hagedorn transseries, we have assumed that $L^{(i)}_{B,F}$ are all positive, and all the terms weighed by complex exponentials of $E$ are implicitly absorbed in $g_{B,F}$. } in $E$:  
\begin{align}
\rho_B(E) &= e^{L^{(1)}_{B} E} f^{(1)}_{B}(E) + e^{L^{(2)}_{B} E} f^{(2)}_{B}(E) +\cdots +g_{B}(E)\\
\rho_F(E) &= e^{L^{(1)}_{F} E} f^{(1)}_{F}(E) + e^{L^{(2)}_{F} E} f^{(2)}_{F}(E) +\cdots +g_{F}(E)
\end{align}
where $L_{B,F}^{(1)} >L_{B,F}^{(2)} > \cdots$ are Hagedorn scales,
 the functions $f_{i}(E)$ have sub-exponential growth at large $E$, 
 \begin{align}
 f_{i}(E) < E^{K_{i}}\exp\left(c_{i} E^{p_{i}}\right)
 \end{align}
with with $p_{i}< 1$ and for some dimensionful parameters 
 $K_{i}, c_{i}$ (with dimensions determined in terms of $\Lambda_{QCD}$ and geometric parameters like volume and curvature).  The functions $g_{B,F}(E)$ are also defined to have sub-exponential growth.

Cancellation of all exponential growth --- which is required for smoothness --- means that 
\begin{align}
\tilde{\rho}(E) = g_B(E) - g_F(E) \, .
\label{eq:naive_cancellation}
\end{align}
Note that this is much weaker than the condition which would be required by supersymmetry, $\tilde{\rho}(E) = 0, E>0$.   But the fact that Eq.~\eqref{eq:naive_cancellation} must hold implies that in adjoint QCD there is a remarkable spectral conspiracy, requiring 
\begin{align}
L^{(i)}_{B} &= L^{(i)}_{F}  \,, \; \forall i \in \mathbb{N} \,, \nonumber \\
f^{(i)}_{B}(E) &= f_{i,F}(E) \,, \; \forall i \in \mathbb{N} \textrm{ and } \forall E>0 \,.
\label{eq:spectral_conspiracy}
\end{align}
By itself this is already surprising and interesting.   Of course, there is a natural follow-up question:  What is the scaling of $\tilde{\rho}(E) = g_{B}(E) - g_{F}(E)$?     

One might naively guess that once Hagedorn cancellations are somehow ensured, $\tilde{\rho} = \rho_{\cal B}(E) - \rho_{\cal F}(E)$ would have the fastest growth allowed by a local quantum field theory in four dimensions.  This guess is motivated by the principle of minimal surprise, since this rate of net growth is all that is necessary for continuity in $L$ at large $N$.  
So one would guess that
\begin{align}
\tilde{\rho}(E) = g_{B}(E) - g_{F}(E) \stackrel{?}{\sim} \exp\left[ (a V)^{\frac{1}{D}} E^{\frac{D-1}{D}} \right]
\end{align} 
where $D=4$ for some  dimensionless parameter $a$ which is roughly the number of degrees of freedom per point.  This growth in the density of states would be associated to a growth in the twisted free energy density of the form $\log \tilde{Z} \sim a V L^{-3}$.  

But as shown by our discussion in the preceding two sections, this guess  is too naive.    The reason is that adjoint QCD on a circle with periodic boundary doesn't just have a smooth dependence on $L$; it actually enjoys large $N$ volume independence.  As we saw in the preceding sections, the resulting cancellations are far stronger than those implied by smoothness. Not even a single 4D particle's worth of density of states remains uncanceled in the $(-1)^F$ graded partition function!  
Consequently, when  4D adjoint QCD lives on a curved compact spatial manifold $M_{3}$, the associated graded density of states behaves as if we were dealing with a two-dimensional theory:     
\begin{align}
\tilde{\rho}(E) \sim \exp\left(\sqrt{b_{\rm eff} \ell E}\right), \;\; \textrm{large} \;  N .
\label{eq:2d_behavior}
\end{align}
Here $\ell = \int_{M_3} d^3{x} \sqrt{g} \, \mathcal{R}$, while the dimensionless parameter $b_{\rm eff}$ is some sort of count of the effective number of degrees of freedom.   The numerical values of this coefficient in large $N$ adjoint QCD as a function of $n_f$ is derived in the Appendix, in a calculable regime.  Reference~\cite{DiPietro:2014bca} showed that $b_{\rm eff} \sim A-C$ in a wide class of SUSY QFTs, where $A$ and $C$ are the 4d conformal anomaly coefficients.\footnote{For discussion of some exceptions to this result see Refs.~\cite{Ardehali:2015bla,DiPietro:2016ond,Hwang:2018riu}.  We note that these exceptions all appear to involve a non-trivial behavior of the color holonomy at small $L$, which is of course also the case in adjoint QCD.}  In adjoint QCD the interpretation of $c_{\rm eff}$ is more mysterious; we have checked that it is not proportional to $A-C$.  However, direct comparison to Ref.~\cite{DiPietro:2014bca} is complicated by the fact that the small $L$ limit of $\tilde{Z}$ and its large $N$ limit do not commute in adjoint QCD except at the supersymmetric point $n_f =1$, because $\log \tilde{Z} \sim (n_f -1)/(N^2L^3)$. 

It is tempting to wonder whether large $N$ 4d adjoint QCD might be related to a 2d quantum field theory, via a relation like  
\begin{align}
\tilde{Z}_{4d}(M_3 \times S^1_L)  = Z_{2d} \, ,
\label{eq:4d_2d}
\end{align}
where on the left we the $(-1)^F$-graded large $N$ partition function of the 4d theory, while on the right $Z_{2d}$ is some (graded) partition function of a 2d quantum field theory.  If such a relation were to hold, then when $M_3 = S^3$, it would be natural to guess that this conjectural 2d QFT should live on a torus with cycle sizes related to $L$ and $\ell$.   The 2d behavior of the density of states $\tilde{Z}$ in Eq.~\eqref{eq:2d_behavior} makes such a conjecture at least conceivable.  Indeed if one sets $M_3 = S^3_R$ and takes the limit $\ell \sim R \ll \Lambda^{-1}$, one can even identify some (chiral) 2d conformal field theories whose partition functions satisfy such a relation\cite{Basar:2015asd}.   While it is interesting that even this much is possible\footnote{The results \cite{Basar:2015asd} were obtained in the free-field limit, and leveraged T-reflection symmetry \cite{Basar:2014mha,McGady:2017rzv,McGady:2018rmo,Duncan:2018wbw,McGady:2018lrz}.  We do not how to generalize the methods of \cite{Basar:2015asd} away from the free-field limit. }, to put such a conjecture on a firmer footing, one would need to make a proposal for what this 2d QFT should be in general. 

 All 4d supersymmetric theories obey Eq. \eqref{eq:2d_behavior}, and appropriately graded partition functions of some supersymmetric theories are known to obey relations like Eq.~\eqref{eq:4d_2d}, see e.g.~\cite{Beem:2013sza,Beem:2014rza,Rastelli:2014jja,Cordova:2017mhb}.  But we are not aware of a concrete 4d-2d connection which would be valid for all 4d supersymmetric theories.  It seems natural to try to understand whether such a generic connection might exist (or perhaps ruled out) for supersymmetric 4d theories, before trying to do understand conjectural 4d-2d connections for confining large $N$ gauge theories like adjoint QCD.

\subsubsection{A comparison of two gradings}
In order to better appreciate the spectral conspiracy that takes place in QCD(adj), i.e. the cancellation of infinitely many Hagedorn growth exponentials of the form $e^{\beta_H E/p}, p=1, 2, \ldots$ in $\tilde Z(L) = \tr e^{-LH} (-1)^F$, it is useful to compare it with a similar-looking construction in a different gauge theory. 
For concreteness, let us consider the  theory on $S^3 \times  S^1$ with a very small $S^3$ radius $R$, $R\Lambda \ll 1$\cite{Unsal:2007fb,Basar:2014jua}. The Hamiltonian is  the one of small-$S^3$ theory.   Its spectrum has both bosonic and fermionic states, and is quantized as $\frac{n}{R}$ and  $\frac{(n+ 1/2)}{R}$ for bosonic and fermionic gauge invariant states in Hilbert space $\cal H= \cal B \oplus  \cal F$, and one can show that\cite{Basar:2014jua}
\begin{align}
\tilde Z(L)_{\rm QCD(Adj)} &= \tr e^{-LH} (-1)^F = \sum_{{\cal B} }e^{- L E_n} {\rm deg}(E_n) -  \sum_{{\cal F} }e^{- L E_n} {\rm deg}(E_n)  \\
&= 1 -4 q^{3/2} + 6 q^2 - 12 q^{5/2} + 28 q^3  - 72 q^{7/2} +168 q^{4} - 364q^{9/2} +828q^{5} \cdots \nonumber
\end{align} 
The non-trivial point established in Ref.~\cite{Basar:2014jua} is that $\tilde Z(L) $ does not have any poles for real positive $L$, meaning that all of the infinitely many Hagedorn poles of the thermal partition function  $Z(\beta)$ disappear as soon as we introduce a grading by the operator  $(-1)^F$ and center symmetry is stable at any $L$  \cite{Unsal:2007fb}. 

The grading by $(-1)^F$ introduces alternating $\pm$ signs for successive energy levels.  One may wonder whether this sort of trick can always cancel off Hagedorn growth.  It turns out that the answer is no!  To see this, consider  pure Yang-Mills theory on $S^3 \times  S^1$. The spectrum of the Hamiltonian is quantized as $\frac{n}{R}$ in the limit $R\Lambda \ll1$, and the confined-phase partition function is known in closed form at large $N$\cite{Aharony:2003sx}.   
 One may ask the following question. If one grades the states in Yang-Mills, similar to $(-1)^F$ grading, by assigning a (+) sign to all $n \in 2\mathbb Z$ (even) states and a (-) sign to all $n \in 2\mathbb Z +1 $ (odd)  states, would one be able to remove all the infinitely many singularities in the confined-phase partition function of large $N$ Yang-Mills? Operationally,  consider 
\begin{align}
\tilde Z(\beta)_{\rm YM} &= \tr e^{-\beta H} e^{i \pi R H}  = \sum_{ n \in  2\mathbb Z }e^{- L E_n} {\rm deg}(E_n) -  \sum_{n \in  2\mathbb Z+1 }e^{- L E_n} {\rm deg}(E_n) \\
&=1+6 q^{2}- 16 q^{3}+72 q^{4} - 240 q^{5}+ \cdots  \nonumber 
\end{align} 
If we were to consider just $Z(\beta) = \tr e^{-\beta H}$, then the   density of states grows as $e^{\beta_H E/p}, p=1, 2, \ldots$ with infinitely many exponentials.  By explicit computation, one can check that grading by $e^{i \pi R H}$ only cancels the leading exponential growth in the set of infinitely many exponential growths.   

Despite the fact that the above construction in Yang-Mills  and QCD(adj) are extremely  similar --- both involve state sums with $(\pm1)$ assigned to interlaced states  --- only in QCD(adj) does one achieve an exact cancellation of the full Hagedorn growth!  This illustrates how special the distribution of states in adjoint QCD is compared to other theories, and is a further piece of evidence for the large $N$ spectral conspiracy of our title.

\subsection{Misaligned supersymmetry in string theory}
What could possibly explain the spectral conspiracy we have observed in adjoint QCD? Standard supersymmetry certainly cannot do the job, and somehow both the large $N$ limit and confinement must play a crucial role in the explanation of the cancellations.  The most satisfying explanation of the cancellations would be directly in field theory and involve some exotic symmetry principle.  But it is hard to find examples of emergent large $N$ symmetries, let alone  ones with the required properties.  The examples we are aware of are the Yangian symmetry of $\mathcal{N}=4$ super-Yang-Mills theory\cite{Dolan:2003uh,Beisert:2017pnr,Beisert:2018zxs} and the large-$N$ spin-flavor symmetry of baryons in QCD with fundamental-representation quarks\cite{Jenkins:1993zu,Dashen:1993as,Dashen:1993jt}.  The Yangian symmetry is (a) tied up with integrability and (b) is a feature of a non-confining theory, so  we see no reason to expect it to have anything to do with our story.  The large $N$ spin-flavor symmetry concerns baryons, while what we need here is something that constraints the glueballs and mesons.   We do not know of any symmetry principles within quantum field theory which depend on both of these features in the necessary way, and looking for such principles is clearly an important topic for future work.  

Something with eerily similar properties to what we need is available in string theory, however.   Without the assumption of spacetime supersymmetry, Kutasov and Seiberg~\cite{Kutasov:1990sv}  showed that in string theories with modular-invariant worldsheet partition functions and no spacetime tachyons, the spacetime density of states graded by $(-1)^F$ has the growth of \emph{at most} a two-dimensional quantum field theory
\begin{align}
\rho_B(E) - \rho_F(E) \lesssim \exp\left(\sqrt{c_{\rm eff} \ell E} \right)
\label{eq:misaligned}
\end{align}
 even when the number of non-compact spacetime directions in the string theory is greater than two.  One might naively guess that this sort of thing could happen due to the emergence of supersymmetry asymptotically, in the sense that the energies of bosonic and fermionic states become degenerate level by level for large energies.  However, Dienes and collaborators pointed out that the mechanism operating in string theory is more subtle.  The cancellations leading to Eq.~\eqref{eq:misaligned} actually come from an oscillatory difference (with an exponential envelope) between the number of bosonic and fermionic states, arranged in such a way that --- in a  sense made precise in \cite{Dienes:1994np} --- the bosonic and fermionic contributions cancel almost exactly, leading to the bound in Eq.~\eqref{eq:misaligned}.\footnote{The cancellations behind misaligned SUSY appear to have the same form as discussed in the preceding section in adjoint QCD, see Ref.~\cite{Dienes:1994np} for a comparison.} This motivated referring to the physics leading to these cancellations as ``misaligned supersymmetry".

Heuristically, this story seems to fit very well with our results, as was noted earlier in Ref.~\cite{Basar:2014jua}.  Large $N$ adjoint QCD should be some sort of free string theory\cite{tHooft:1973alw,Witten:1979kh}, and this string theory should be well-defined, without tachyonic modes in spacetime.  Could it be that adjoint QCD furnishes the first QFT example of the string-theoretic ``misaligned supersymmetry" idea of Kutasov, Seiberg, and Dienes?

This is a tantalizing possibility, but it is not easy to make it precise.  Even with supersymmetry, the worldsheet descriptions of string duals to large $N$ gauge theories are subtle\cite{Berkovits:1999im,Metsaev:1998it,Berkovits:2000fe}, because the associated dual gravity backgrounds involve Ramond-Ramond (RR) flux\cite{Maldacena:1997re}.  The known constructions for the dual of $\mathcal{N}=4$ SYM leads to a conformal world-sheet sigma model\cite{Berkovits:2000fe}, so the associated worldsheet partition function must be modular-invariant.   When the field theory lives on $\mathbb{R}^4$,  applying the technology of Refs.~\cite{Kutasov:1990sv,Dienes:1994np,Dienes:1994jt,Dienes:1995pm} to this model is guaranteed to give a trivial result due to $\mathcal{N}=4$ supersymmetry.   But when the boundary geometry is $S^3 \times S^1$, and fermions have periodic boundary conditions on $S^1$, the field theory is guaranteed to be in the confining phase for all $S^1$ sizes $L$\cite{Witten:1998qj}.  The associated field theory partition function is then non-trivial and must obey Eq.~\eqref{eq:SUSYscaling}.  It is then natural to expect that the dual string theory spectrum non-trivially satisfies the misaligned supersymmetry constraints.  It would be interesting to check this explicitly for the string dual of $\mathcal{N}=4$ SYM using the worldsheet CFT proposed in Ref.~\cite{Berkovits:2000fe}.

A much more serious check of the relation to misaligned supersymmetry would involve finding a non-supersymmetric string theory living in a bulk with RR flux, and showing that it is associated to a local conformal sigma model on the worldsheet, so that it has a modular-invariant worldsheet partition function.  If this can be done, one could presumably use the results of Refs.~\cite{Kutasov:1990sv,Dienes:1994jt,Dienes:1994np,Dienes:1995pm} to obtain a non-trivial constraint of the form of Eq.~\eqref{eq:misaligned} on the string spectrum in the bulk, and then use the AdS/CFT dictionary to translate these constraints to a statement about the dual gauge theory.  Checking whether this works in any explicit example is, of course, very difficult.  Perhaps more importantly, there is also a conceptual challenge: it is not clear how, in general, confinement in the field theory is supposed to interact with misaligned supersymmetry on the string theory side of the story.  Yet confinement plays a crucial role in our field theory arguments.  To sharpen the claim that the Bose-Fermi cancellations seen in adjoint QCD are tied up with misaligned supersymmetry, we need some way to fill this conceptual gap in the argument. 

\subsection{Connection to QCD}

One may wonder whether the story in this paper can be brought closer to real QCD.  We can make two remarks concerning this question. First, throughout this paper we have focused on $U(N)$ adjoint QCD, but in QCD the gauge group is $SU(N)$. So what happens to our story in $SU(N)$ adjoint QCD?  In the large $N$ limit the parallel of Eq.~\eqref{eq:UNresult} is
\begin{align}
\log \tilde{Z}(L) \sim -\frac{\frac{\pi^2}{45} (n_f-1) V}{L^3} + b \frac{\ell}{L} + \cdots \,,\; \textrm{$SU(N)$ adjoint QCD.} 
\label{eq:SUNresult}
\end{align}
It should be emphasized that  at large $N$ the coefficient of $L^{-3}$ in this expression is exactly determined in free field theory. This can be contrasted with the situation in thermal Yang-Mills theory, where the coefficient of $L^{-3}$ involves a non-trivial series in $\lambda$.
The reason that the coefficient of $L^{-3}$ is determined in free field theory in $SU(N)$ adjoint QCD is that the difference between $U(N)$ and $SU(N)$ adjoint QCD is just the addition of a free Maxwell field and $n_f$ free Majorana fermion fields.

Second, one can ask whether our results about adjoint QCD might have any bearing about on QCD as it is seen in nature --- at least to the extent that the large $N$ limit is useful in QCD.    Naively it seems like there cannot be any connection, because real QCD has fundamental fermions rather than adjoint fermions.   But at $N=3$ the fundamental (F) representation is isomorphic with the two-index anti-symmetric (AS) representation Dirac fermions.  This means that one can keep the fermions in either the F or in the AS representations when taking the large $N$ limit.  Both limits have reasonable (but distinct) phenomenologies compared to $N=3$ expectations
\cite{Armoni:2003yv,Armoni:2004uu,Cherman:2009fh,Cherman:2012eg,Armoni:2014ywa}.  

The relevance of these comments is that in the large $N$ limit there is a precise relation between adjoint QCD and QCD(AS):  the correlation functions of local charge-conjugation-even bosonic operators in these theories coincide up to $1/N$ corrections thanks to ``orientifold large $N$ equivalence" \cite{Armoni:2004uu,Armoni:2003gp,Armoni:2004ub}.  While the cancellations we have discussed in adjoint QCD are between bosonic and fermionic states, it seems likely that any explanation of the cancellations would lead to strong constraints on the bosonic states.  If this is the case, then the large $N$ equivalence described above would lead to constraints on the large $N$ spectrum of QCD(AS), and hence teach us about an unconventional, but reasonable, large $N$ limit of real QCD.

\subsection{Implications for the vacuum energy}
\label{sec:CC}

We now explain what our results say about the vacuum energy $\langle E \rangle$ of adjoint QCD.\footnote{Our discussion is conceptually very similar to the work on supertrace  relations the context of misaligned SUSY in string theory in Refs.~\cite{Dienes:1994jt,Dienes:1994np,Dienes:1995pm,Dienes:2001se}.  In field theory, Refs.~\cite{Cherman:2015rpc,Basar:2014hda,Basar:2014mha,Basar:2015asd,Basar:2014jua,Basar:2015xda} discussed one-loop evidence for constraints on the vacuum energy in a class of theories which includes adjoint QCD.  Our results here generalize and simplify this earlier discussion. }
The vacuum energy is generically is a scheme-dependent UV-sensitive quantity.  We will evaluate it using several natural UV regulators, which all give the same results.  We view this as evidence that our conclusions are physically significant.

To get started, it turns out to be easiest to compute $\langle E \rangle$ using a spectral heat kernel regulator, which introduces a damping factor $\frac{1}{\mathcal{N}(\mu)} e^{-\omega_n/\mu}$ into the spectral sum over the energies $\omega_n$.  Here $\mathcal{N}(\mu) = \sum_n e^{-\omega_n/\mu}$ is a normalization factor and $\mu$ is the effective cutoff scale.  We are interested in computing the heat-kernel regularized sum for the vacuum energy
\begin{align}
\langle E(\mu) \rangle =  \frac{1}{\mathcal{N}(\mu)} \sum_{n} (-1)^F \omega_n e^{-\omega_n/\mu} \,,  \qquad \textrm{heat kernel.}
\end{align}
Here we have assumed that the spectrum is discrete, as would be the case in any finite spatial volume $V$. In a generic 4D QFT, evaluating such a sum for large cutoff $\mu$ leads to 
$\langle E(\mu) \rangle\sim V \mu^4 $.  But by identifying $\mu = 1/L$ one can note that this expression coincides with $\partial_{L} \log \tilde{Z}(L)$, where $\tilde{Z}$ is the partition function on a circle with periodic boundary conditions.   The preceding sections then imply that in large $N$ adjoint QCD 
 \begin{align}
\langle E(\mu \gg \Lambda) \rangle \sim \Lambda_{QCD}^4 V \,,  \qquad \textrm{heat kernel.}
\label{eq:vacuum_energy}
\end{align}
Note that there is no explicit dependence on the UV cutoff $\mu$ in Eq.~\eqref{eq:vacuum_energy}.  This result follows from the structure of Eq.~\eqref{eq:general_structure_Z} and Eq.~\eqref{eq:SUSYscaling}, as well as the observation that the term proportional to $L$ in Eq.~\eqref{eq:general_structure_Z}  is given by the spatial integral of the gluon condensate $\langle \tr F^2 \rangle \sim \Lambda^4$.  But $\Lambda \sim \mu \exp[-8\pi^2/(\beta_0 \lambda)]$, where $\beta_0 = 11/3 - 2/3 n_f$ is the one-loop beta function coefficient and $\lambda = \lambda(\mu)$.   Therefore $\langle E \rangle$ is exponentially small compared to the UV cutoff $\mu$ for large $\mu$.

This is a rather provocative statement, so it is important to understand it better.  Indeed, one might be concerned that, because we are dealing with a scheme dependent quantity, maybe we happened to pick a regulator that somehow automatically makes the coefficient of $\mu^4$ small.  But in fact the result is driven by the physical spectral properties of adjoint QCD, and so we would expect it to hold with any reasonable choice of regulator.  

We illustrate this by showing how the calculation works with other regulators.  First, we consider $\langle E \rangle$ evaluated with a hard-cut-off regulator.  Hard cut-off regulators are not compatible with gauge invariance when used at the level of quarks and gluons, but here we envision using such a regulator to compute the  contributions of the physical color-singlet particle excitations to  $\langle E\, \rangle$. With this in mind, we write 
\begin{align}
\langle E(\mu)\rangle = \int^{\mu}_{0} d E \,\tilde{\rho}_C(E) E\,, \qquad \textrm{hard cut-off}
\label{eq:vacuum_energy_hard_cutoff}
\end{align}
Here $\tilde{\rho}_C(E)$ is the $(-1)^F$-graded canonical (that is, single-particle) density of states, which is to be distinguished from the grand-canonical density of states $\tilde{\rho}(E)$ we have been discussing in most of the paper.   In finite volume $\tilde{\rho}(E)$ and $\tilde{\rho}_C(E)$ are given by sums of delta functions, but for large $E$ it becomes meaningful to view them as smooth functions of $E$.    There is a simple relation between the large-energy behavior of the grand-canonical and canonical densities of state:
\begin{align}
\tilde{\rho}(E) \sim \exp \left[ \left(x E\right)^{\frac{d-1}{d}} \right]  \qquad  \Longleftrightarrow  \qquad \tilde{\rho}_C(E) \sim (x E)^{d-2}
\label{eq:density_dictionary}
\end{align}
Here $x$ is a parameter with dimensions of length, and $d$ is an effective spacetime dimension.  In large $N$ adjoint QCD on a flat spatial manifold, we have seen that the small-$L$ expansion of $\log \tilde{Z}(L)$ starts with a term linear in $L$.  This maps to taking $d=0$ in Eq.~\eqref{eq:density_dictionary}, so that there are  no power divergences in $\langle E(\mu)\rangle$ with a hard cut off once one takes into account the sum over particle species at large $N$.  The largest growth allowed is
\begin{align}
\tilde{\rho}_C(E)  \sim \Lambda^4 V E^{-2}\,, \qquad \textrm{large $N$ adjoint QCD} \,.
\end{align}
Plugging this into Eq.~\eqref{eq:vacuum_energy_hard_cutoff} we land on the same result as in Eq.~\eqref{eq:vacuum_energy}.

For a third example, consider adjoint QCD with $n_f \le 4$.  Then we can regularize $U(N)$ adjoint QCD by embedding it into $U(N)$ $\mathcal{N}=4$ SYM theory\cite{ArkaniHamed:1997mj}, which is a UV-finite theory.   One can make the regularized theory flow to adjoint QCD by turning on SUSY-breaking mass terms for the adjoint scalars and some of the adjoint fermions.   Note that this regulator respects all of the symmetries of adjoint QCD, including center symmetry.  

For simplicity, we focus on $n_f = 4$ adjoint QCD, and give all six adjoint scalars a common mass $m_s$.  Choosing the bare value of $\lambda$ to be small guarantees that it will be small at the scale $m_s$, where it starts running, and consequently   $m_s \gg \Lambda$.  How does the vacuum energy depend on the regulator scale $m_s$ with this more elaborate regularization scheme?  To answer this question, we first observe that $\mathcal{N}=4$ SYM theory enjoys unbroken center symmetry even after we break supersymmetry by taking $m_s \neq 0$.   To see this recall that when $m_s =0$ the GPY holonomy effective potential vanishes both perturbatively and non-perturbatively;  this is because a superpotential is forbidden by $\mathcal{N}=4$ supersymmetry.  But when all six scalars are given a mass $m_s$, there is a one-loop contribution to the holonomy effective potential:
\begin{align}
V_{\rm eff}(\Omega) = \frac{2}{\pi^2L^4} \sum_{n\ge 1} \frac{|\tr\Omega^n|^2}{n^4}  \left[3 - 3(n L m_s)^2 K_2(n L m_s) \right]\,,
\label{eq:GPY_N4}
\end{align}
This potential implies that center symmetry is not spontaneously broken at small $L \Lambda$ for \emph{any} value of $m_s \ge 0$.  With this in mind, we can again consider the vacuum energy in the regularized theory
\begin{align}
\langle E(m_s)\rangle = \int^{\infty}_{0} d E \,\tilde{\rho}_C(E; m_s) E\,, \qquad \textrm{$\mathcal{N}=4$ SYM regulator.}
\end{align}
Here $\tilde{\rho}(E; m_s)$ is the regularized canonical graded density of states.   When $E \gg m_s$, standard $\mathcal{N}=4$ supersymmetry implies that $\rho(E \gg m_s) = 0$.  But when $\Lambda \ll E \ll m_s$, our remarks concerning center symmetry above imply that the density of states of softly-broken $\mathcal{N}=4$ SYM theory enjoys the same large $N$ spectral cancellations that we have discussed in the preceding sections.  As a result, $\tilde{\rho}(E;m_s)$ scales as Eq.~\eqref{eq:density_dictionary} with $d=0$ when $\Lambda \ll E \ll m_s$!  This implies that we again land on Eq.~\eqref{eq:vacuum_energy} with the $\mathcal{N}=4$ regulator, just as we did with the other regulators we have considered.

Now that we have seen that $\langle E \rangle$ is very small in adjoint QCD at large $N$, and is given by the $\mathbb{R}^4$ limit of $\langle \tr F^2 \rangle$, it is time to ask about its sign.  The operator $\tr F^2$ is non-negative, and if the same is true for the path integral measure, then $\langle E \rangle$ must also be non-negative.  Indeed, when $n_f$ is even, one can package the fermions as Dirac spinors.  Integrating out the fermions gives $(\det D)^{n_f/2}$, where $D$ is the Dirac operator, whose non-zero eigenvalues come in conjugate pairs thanks to $\gamma_5$ hermiticity.  Therefore $(\det D)^{n_f/2}$ is non-negative for even $n_f$.  When $n_f$ is odd, integrating out the fermions gives a Pfaffian, which is the square root of the determinant up to a sign.  To see that this sign can be consistently chosen to be $+1$, note that one can add a gauge-invariant positive mass term for the fermions, which eliminates all zero modes of the Dirac operator, while keeping the determinant positive.  Then we can define the Pfaffian to be positive for some reference field configuration, say $A_{\mu} = 0$, and ask whether it can change sign as we vary $A_{\mu}$.  But at finite positive $m$ this is impossible since there are no zero modes.  Therefore the Pfaffian can consistently be defined to be positive for any finite positive $m$.  The same must therefore be true as we take the $m = 0$ limit with $m \in \mathbb{R}^{+}$.   So for any $n_f$ we find that 
\begin{align}
\langle E(\mu) \rangle \ge 0
\label{eq:positive_E}
\end{align}
 in massless adjoint QCD at large $N$.

The inequality in Eq.~\eqref{eq:positive_E} is saturated at $N_f =1$ due to supersymmetry.  It is also saturated  when $n_f$ is within the conformal window, because one-point functions in a conformal field theory  on $\mathbb{R}^4$ must vanish, and large $N$ volume independence implies that this is also true on $\mathbb{R}^3 \times S^1$.  For $n_f = 2,3$, where the theory is likely not conformal in the infrared, the vacuum energy must be positive, and exponentially small compared to the UV cutoff.

To summarize, we have presented evidence that there exists a family of non-supersymmetric quantum field theories --- large $N$ $U(N)$ adjoint QCD --- whose vacuum energy is non-negative and exponentially small.  We are aware of only two sets of solid examples of this sort of behavior in the field theory literature, and another set in the string theory literature.  The first field theory example is Witten's result concerning $2+1$ dimensional supergravity\cite{Witten:1994cga}.  Witten pointed out that, in this setting, supersymmetry can be unbroken, ensuring that the vacuum energy vanishes exactly, without ensuring Bose-Fermi pairing among the excited states.  This has a clear surface-level resemblance to our story, but we do not know how to make the connection deeper.  The second set of field-theory examples involves microscopically-massless theories with spontaneously broken supersymmetry. In that context it is famously the case that $\langle E \rangle > 0$, while the fact that $\langle E \rangle$ is exponentially small compared to a UV cutoff follows from dimensional transmutation\cite{Iliopoulos:1974zv,Witten:1981nf}.   Finally, Refs.~\cite{Dienes:1994np,Dienes:1995pm} showed that a vanishing vacuum energy can appear in non-supersymmetric perturbative string theory as a consequence of misaligned SUSY.  Our story is distinguished from all of these examples by being established in a manifestly non-supersymmetric quantum field theory, whose spectrum certainly does not feature level-by-level Bose-Fermi pairing, and we have not made any appeal to string theory (except perhaps indirectly, by taking a large $N$ limit) to establish our results.

Given how few ways of getting a small vacuum energy are known in quantum field theory, it would be interesting to explore adjoint QCD and its large $N$ limit more deeply, with the goal of developing some symmetry-based explanation for the cancellations we have found.  Perhaps this exploration can also inspire some eventual phenomenological applications.  


\section*{Acknowledgments}

We are very grateful to Keith Dienes, Patrick Draper, Zohar Komargodski, and David McGady for helpful comments, and are especially indebted to Larry Yaffe for a crucial suggestion on our analysis and extensive discussion concerning holonomy effective potentials.  This work was supported in part by the DOE grant DE-sc0011842 and by the National Science Foundation under Grant No. NSF PHY17-48958, and we are grateful to Kavli Institute for Theoretical Physics for its hospitality during the early stages of this work.  M. \"U. acknowledges support from U.S. Department of Energy, Office of Science, Office of Nuclear Physics under Award Number DE-FG02-03ER41260.

\appendix
\section{Coefficient of $L^{-1} \int_{M_3} d^{3}x \, \sqrt{g} \,\mathcal{R}$}
\label{sec:appendix}
In this appendix we compute the coefficient of $L^{-1} \int_{M_3} d^{3}x \, \sqrt{g} \,\mathcal{R}$ in the small-$L$ expansion of $\log \tilde{Z}$ in adjoint QCD at large $N$.  In supersymmetric theories, this coefficient was related to conformal anomaly coefficients in Ref.~\cite{DiPietro:2016ond}, so that it counts the number of degrees of freedom of the theory.   Here we briefly discuss the extent to the coefficient of $L^{-1}$ in adjoint QCD counts the effective number of degrees of freedom in large $N$ adjoint QCD.\footnote{We are grateful to Zohar Komargodski for asking us this question, which prompted this appendix.}  We do this when $M_3 = S^3$, with radius $R$, and assume that $R\Lambda \ll 1$.  This allows us to ignore everything except one-loop effects when doing the calculation.   In large $N$ adjoint QCD
\begin{align}
\log \tilde{Z} = b \frac{R}{L} + \cdots
\end{align}
and our goal here is to compute $b$.  This determines the coefficient of $L^{-1} \int_{M_3} d^{3}x \, \sqrt{g} \,\mathcal{R}$, since 
\begin{align}
L^{-1} \int_{M_3} d^{3}x \, \sqrt{g} \,\mathcal{R}  =\big|_{M_3 = S^3} = \frac{R}{L} 12\pi^2
\end{align}
given that $\mathcal{R} = 6/R^2$ on $S^3$. 

It might be tempting to compute $b$ directly from the holonomy effective potential, by writing
\begin{align}
L V_{\rm eff}(\Omega) = \frac{1}{L^3} \sum_{n \ge 1} c_n(L/R) |\tr \Omega^n|^2
\end{align}
expanding $c_n(L) = c_n^{(0)} +  c_n^{(1)} L^2/R^2$, and looking at the coefficient of $R/L$ in the resulting expression when $\Omega$ is set to its center-preserving value.    This is not quite correct due to the non-commutativity of the large $N$ and small $L$ limits.  We need to take the large $N$ limit first, compute the partition function --- which entails taking into account fluctuations around the confining saddle-point of the path integral --- and only then extract the desired coefficient.   As explained in Refs.~\cite{Basar:2014jua,Basar:2015asd}  the $S^{3} \times S^{1}$ graded partition function of adjoint QCD takes the form
\begin{align}
\tilde{Z} &= \prod_{n=1}^{\infty} \frac{(1-Q^{2n})^3}{1 - 3Q^{2n} +4 n_f Q^{3n} - 3Q^{4n}+ Q^{6n}} 
\end{align}
where $Q = e^{-\beta/(2R)}$.  To extract $b$, it is very useful to rewrite this expression in terms of standard modular functions
\begin{align}
\tilde{Z}(\tau) = 8 \eta(\tau)^{9} \prod_{\alpha = 1}^{3} \left[ e^{-i \pi b_{\alpha}} \cos(\pi b_{\alpha}) \, \frac{1}{ \Th{1/2}{b_{\alpha}}(\tau)}
\frac{1}{\Th{0}{b_{\alpha}}(\tau) } \right] \,, 
\label{eq:modular}
\end{align}  
where $\tau$ is defined via $L/R = 2\pi i \tau$, $\eta$ is the Dedekind eta function, and our conventions for the theta functions are the same as those of Ref.~\cite{Basar:2015asd}. The numerical parameters $b_{\alpha}$, written in the form  $z_{\alpha} = e^{2\pi i b_{\alpha}}$, are are given by
\begin{align}
z_1&=\frac{\kappa ^2+2-\sqrt{\kappa^4+4}}{2 \kappa} \nonumber \\
z_2&= -{1 \over 16\kappa^2} \left[ \kappa^3+2\kappa -2\sqrt{\eta} +\left((\kappa^3+2\kappa-2\sqrt{\eta})^2-16\kappa^4\right)^{1/2}\right] \nonumber \\
z_3&=-{1 \over 16 \kappa^2} \left[ \kappa^3+2\kappa +2\sqrt{\eta} -\left((\kappa^3+2\kappa+2\sqrt{\eta})^2-16\kappa^4\right)^{1/2}\right]
\end{align}
where
\begin{align}
\kappa &= \left(2n_f+2\sqrt{n_f^2-2}\right)^{1/3}\\
\eta &= 3\left( \kappa^4-n_f \kappa^3- \kappa^2+  2 \right)\,.
\end{align}

\begin{figure}[th]
\centering
\includegraphics[width=.6\textwidth]{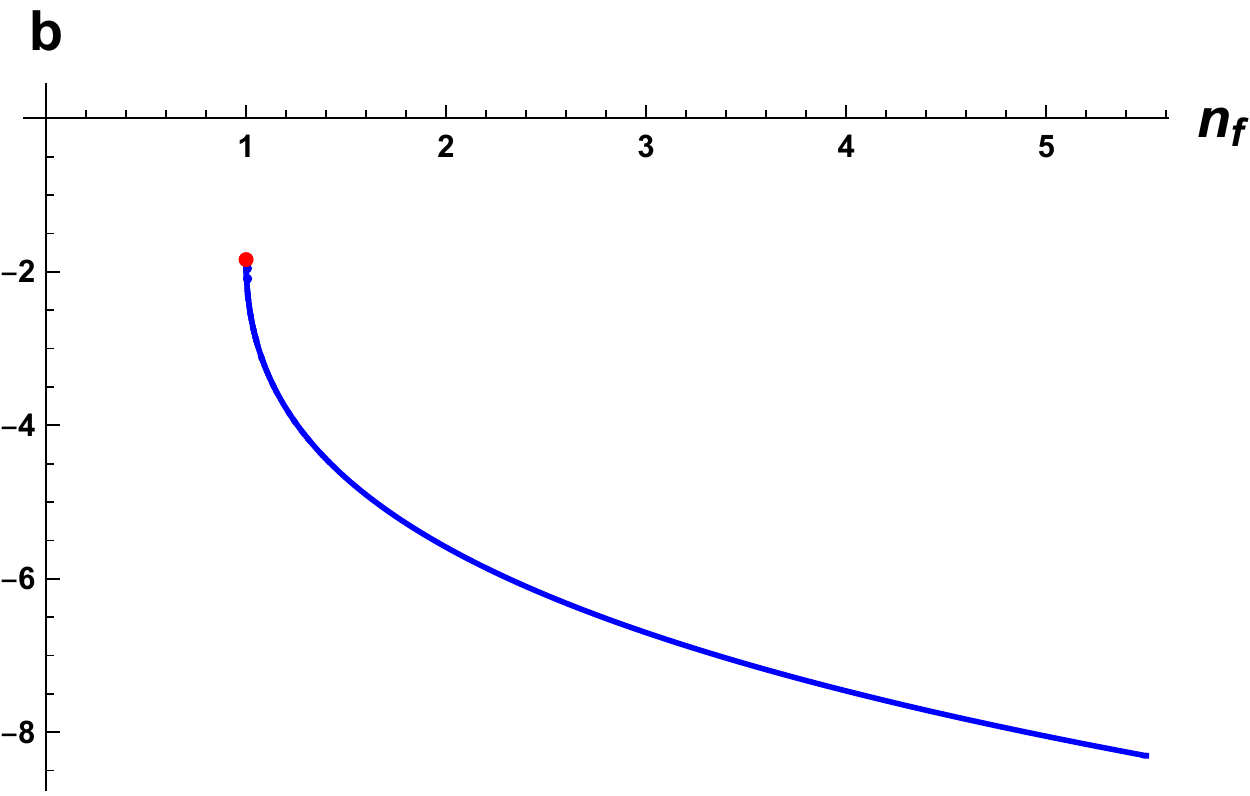}
\caption
    {%
 The coefficient $b$ of $1/\beta$ in the small-$\beta$ expansion of $\log \tilde{Z}$ as a function of $n_f$ in adjoint QCD on $S^3_{R} \times S^1_{\beta}$, calculated assuming that the $S^3$ radius $R$ is small, $R\Lambda \ll 1$.  The value of $b$ at $n_f=1$ is plotted in red, and occurs at a cusp of the function $b(n_f)$.
    }
\label{fig:bPlot}
\end{figure}

The reason Eq.~\eqref{eq:modular} is useful is that it allows us to use the modular S transformation properties of the $\eta$ and $\vartheta$ functions to compute the small-$|\tau|$ behavior of $\log \tilde{Z}$.  Some algebra yields
\begin{align}
b = - \frac{3\pi^2}{2} + 4 \pi^2\sum_{a = 1}^{3} b_{\alpha}^2 \, .
\end{align} 
It can be shown that the function $b = b(n_f)$ is smooth for $n_f >1$, but has a cusp at $n_f = 1$, as shown in Fig.~\ref{fig:bPlot}.~\footnote{Defining $b$ below $n_f = 1$ is much more subtle. It is discussed in Refs.~\cite{Basar:2015xda,Basar:2015asd}.}       Of course, physically, the parameter $n_f$ is an integer, so $n_f$ takes discrete values as a function of $n_f$, given in Table~\ref{table:bValues}.

\begin{table}
\begin{center}
  \begin{tabular}{c|c|c|c|c|c}
    $n_f$ & $1$ & $2$ & $3$ & $4$ & $5$ \\ \hline
     b & $-1.84$ & $-5.59$ & $-6.70$ & $-7.46$ & $-8.05$ 
 \end{tabular}
 \caption{Values of $b$ at various values of $n_f$, rounded to two decimal places.}
  \label{table:bValues}
 \end{center}
\end{table}

  As a more physical way of effectively varying $n_f$, we have also checked that if one fixes e.g. $n_f = 3$, and keeps  two quark flavors massless while varying the mass $m$ of the remaining flavor, $|b|$ decreases with increasing $m$.    This suggests that it may be possible to interpret $|b|$ as a counter of the number of effective ``degrees of freedom" of large $N$ adjoint QCD.  It would be very interesting to make this interpretation more precise.

\newpage
\bibliography{small_circle}
\bibliographystyle{JHEP}

\end{document}